\documentclass[preprint]{JASAnew}
\usepackage{tipa} 
\usepackage{booktabs} 

\begin{document}

\title[GESI for predicting Speech Intelligibility]{GESI: Gammachirp Envelope Similarity Index for Predicting Intelligibility of Simulated Hearing Loss Sounds}
\author{Ayako Yamamoto}\email{yamamoto.ayako@g.wakayama-u.jp}
\affiliation{Graduate School / Faculty of Systems Engineering, Wakayama University, Sakaedani 930, Wakayama, Wakayama 640--8510, Japan}

\author{Toshio Irino}\email{irino@wakayama-u.ac.jp}
\affiliation{Graduate School / Faculty of Systems Engineering, Wakayama University, Sakaedani 930, Wakayama, Wakayama 640--8510, Japan}

\author{Fuki Miyazaki}\email{miyazaki.fuki@g.wakayama-u.jp}
\affiliation{Graduate School / Faculty of Systems Engineering, Wakayama University, Sakaedani 930, Wakayama, Wakayama 640--8510, Japan}

\author{Honoka Tamaru} 
\affiliation{Graduate School / Faculty of Systems Engineering, Wakayama University, Sakaedani 930, Wakayama, Wakayama 640--8510, Japan}

 


\date{\today} 

\begin{abstract}
\nolinenumbers

We propose an objective intelligibility measure (OIM), called the Gammachirp Envelope Similarity Index (GESI), which can predict the speech intelligibility (SI) of simulated hearing loss (HL) sounds for normal hearing (NH) listeners. GESI is an intrusive method that computes the SI metric using the gammachirp filterbank (GCFB), the modulation filterbank, and the extended cosine similarity measure. The unique features of GESI are that i) it reflects the hearing impaired (HI) listener's HL that appears in the audiogram and is caused by active and passive cochlear dysfunction, ii) it provides a single goodness metric, as in the widely used STOI and ESTOI, that can be used immediately to evaluate SE algorithms, and iii) it provides a simple control parameter to accept the level asymmetry of the reference and test sounds and to deal with individual listening conditions and environments. We evaluated GESI and the conventional OIMs, STOI, ESTOI, MBSTOI, and HASPI versions 1 and 2 by using four SI experiments on words of male and female speech sounds in both laboratory and remote environments. GESI was shown to outperform the other OIMs in the evaluations. GESI could be used to improve SE algorithms in assistive listening devices for individual HI listeners.

\end{abstract}



\maketitle


\nolinenumbers





\section{\label{sec:1} Introduction}

It is now critical to develop the next generation of assistive listening devices that can compensate for the hearing difficulties of individual hearing impaired (HI) listeners. Speech enhancement and noise reduction algorithms~\citep{loizou2013speech} should become more robust and effective based on the individual hearing characteristics. 
For this purpose, subjective listening tests to measure individual speech intelligibility (SI) are essential, but these tests are time consuming and costly. It is also important to develop an effective objective intelligibility measure (OIM) that can predict the SI of HI listeners whose hearing levels are individually different.

Many OIMs have been proposed to evaluate speech enhancement (SE) and noise reduction algorithms for improving SI \citep{falk2015objective, van2018evaluation}. 
STOI~\citep{taal2011algorithm} and ESTOI~\citep{jensen2016algorithm} have been the most popular OIMs for this purpose. \cite{yamamoto2020gedi} also proposed GEDI (Gammachirp Envelope Distortion Index, [\textdyoghlig\'eda\i]) for more conservative evaluation.
These are intrusive methods and do a good job of predicting the SI for normal hearing (NH) listeners.
They provide a single metric value that can be converted to the SI values of phonemes and words with a monotonic sigmoid function. 
Therefore, it was assumed that the metric would be used directly as a measure of goodness when comparing the proposed and conventional SE algorithms. 
Although the realistic SI values may differ by individual speech-in-noise conditions, the metrics are as convenient as commonly used metrics (e.g., SNR and SDR) for evaluating the SE algorithms.
However, they cannot assess the SI of HI listeners whose hearing levels are elevated. It is important to address this issue in the development of SE algorithms for the assistive listening devices for HI listeners.

There have been several approaches for this.
\cite{kates2014hearing} proposed HASPI (Hearing-Aid Speech Perception Index version 1, HASPIv1) to reflect the audiograms of HI listeners although it has rarely been used in algorithm development. They also proposed HASPI version 2 (HASPIv2) to improve the SI prediction of sentences~\citep{kates2021hearing}.
More recently, \cite{irino2022speech} have proposed an OIM called GESI (Gammachirp Envelope Similarity Index, [\textdyoghlig\'esi]) by extending the algorithm in GEDI~\citep{yamamoto2020gedi}.
These methods can reflect the hearing level and the degradation factors of the peripheral active mechanism of an HI listener.
These OIMs are based on spectral analysis, feature extraction, and correlation or similarity between the test and reference signals. Except for the neural network (NN) output of HASPIv2, the metric is extracted as bottom-up features analyzed only from speech sounds.

An alternative method is to use training data to infer the SI value of HI listeners, such as the NN output of HASPIv2, which was trained using SI data from the HINT database~\citep{nilsson1994development}. More recently, deep neural networks (DNNs) have been introduced to improve the OIMs. For example, in the first Clarity Prediction Challenge (CPC1) ~\citep{barker20221st} for the SI prediction of HI listeners, many systems (e.g.~\citealt{huckvale2022elo, tu2022exploiting, kamo2022conformer})
outperformed the baseline system, which relies on the combination of 
MBSTOI~\citep{andersen2018refinement} and a HL simulator developed in Cambridge University (CamHLS, hereafter; \citealp{baer1993effects, nejime1997simulation, stone1999tolerable}). 
In the recent Clarity Challenge workshop~\citep{claritychallenge}, even non-intrusive OIM performs well in SI prediction for HI listeners.
This high performance was made possible by using a huge amount of training data to train several megabytes of parameters for prediction. In addition, it is possible to estimate a complex mapping function between the low-level spectral features and the SI values, which may also be influenced by cognitive factors of the HI listener. 

However, it is well known that the range of good performance is usually limited by the prepared training data and the computational power and memory. For the evaluation of SE algorithms, it is necessary to collect all possible (or at least massive) speech and noise sounds that may be encountered in everyday life, and the corresponding responses of individual HI listeners, whose characteristics may vary tremendously.
Therefore, such a data-driven approach is very complex and does not appear to be ready for immediate use in SE algorithm development.
In addition, these OIMs have not provided a method to compensate for SI predictions that depend on listening conditions such as ambient noise level and audio device quality.

Therefore, it is still desirable to use a simple OIM that does not rely on training data and provides a good metric, highly correlated with SI, from the input signals alone. Furthermore, the OIMs could be still improved by introducing psychophysical and physiological knowledge of the auditory system. Then it would be possible to analyze the degradation factors to see if the degradation of the SI is caused by the peripheral HL of more central factors. Such OIMs can also be used as a front-end to the DNN methods, which could improve the performance compared to the end-to-end (i.e., signal-to-SI) DNN methods, at least when the variation and amount of the training data is limited.

GESI was developed with this in mind~\citep{irino2022speech}. GESI can reflect the HL of the HI listener that appears on the audiogram and is caused by active and passive cochlear dysfunction. In the previous study~\citep{irino2022speech}, it was shown that GESI is better than STOI, ESTOI, and HASPIv1. However, the evaluation was limited to predicting the average SI of male speech using simulated HL sounds. It is important to predict SI without using simulated HL sounds because HI listeners hear normal speech sounds, not simulated HL sounds. In addition, several questions remain unanswered. \textit{i)} Is SI prediction of female speech possible? GESI uses fundamental frequency $F_o$ information, which was set to an average male $F_o$ in the previous version. It should be extended to reflect the SI values of female speech based on additional subjective experiments. \textit{ii)} Is it possible to predict SI values for individual listeners instead of an average? The SI can change depending on listening conditions, such as the level of ambient noise and the audio device.
\textit{iii)} Is GESI better than the newer and more sophisticated version of HASPI, i.e., HASPIv2?





In this study, to answer these questions, 
we have expanded the range of SI experiments and improved the algorithm to take into account the experimental results.
First, in section~\ref{sec:OIM}, we explain the algorithm of GESI and summarize the conventional OIMs, i.e., STOI, ESTOI, MBSTOI, HASPIv1, and HAPSIv2, used in the comparison. In section~\ref{sec:SubjectiveExp}, we explain the subjective SI experiments conducted in laboratory and crowdsourced remote environments, after describing the motivation for such experiments. Finally, in section~\ref{sec:EvaluationOIM}, GESI was evaluated and compared with the conventional OIMs based on the experimental results, in three evaluation schemes.

\section{Proposed and conventional OIMs}
\label{sec:OIM}

We describe algorithms of GESI and the conventional OIMs, STOI, ESTOI, MBSTOI, HASPIv1, and HASPIv2 for comparison.

\subsection{Algorithm of GESI}
\label{sec:AlgorithmGESI}

\begin{figure}[t]
\centering
\includegraphics[width=1\linewidth]{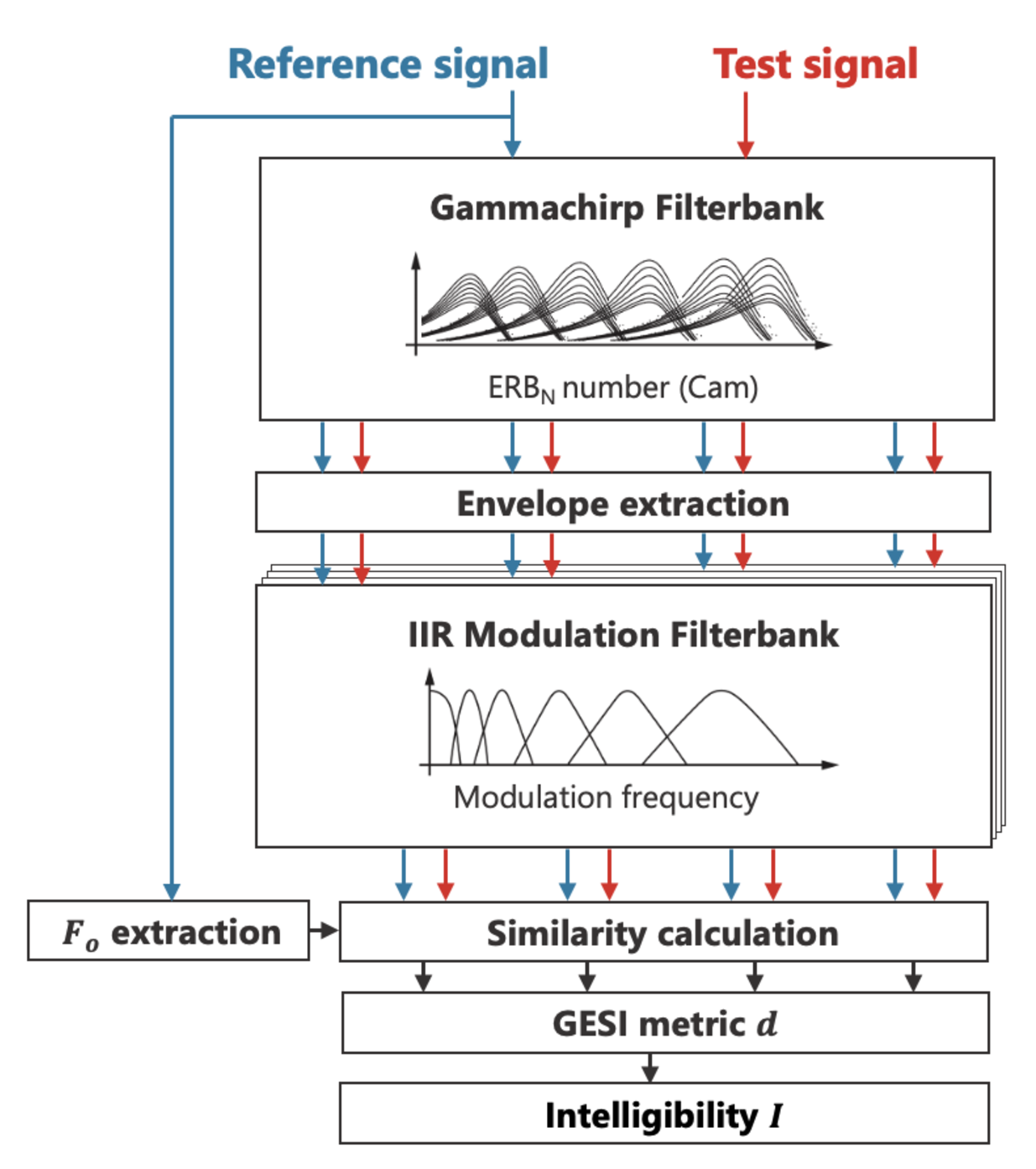}
  \caption{ \label{fig:GESI_Blockdiagram} Block diagram of GESI
} 
\end{figure}

We have developed GESI based on a framework similar to the Gammachirp Envelope Distortion Index (GEDI, [\textdyoghlig\'eda\i]) \citep{yamamoto2020gedi}. Figure~\ref{fig:GESI_Blockdiagram} shows a block diagram of GESI.
The input sounds to GESI are reference and test signals. The process starts with a frame-based version of the dynamic compressive gammachirp filterbank (GCFB; \citealp{irino2006dynamic, irino2023hearing}).
 The hearing level and a compression health parameter of an HI (and NH) listener can be set to simulate the auditory peripheral characteristics. 
The next steps are envelope extraction and filtering with an IIR version of a modulation filterbank (MFB), first introduced in sEPSM \citep{jorgensen2011predicting, jorgensen2013multi}.

Then, we apply a new method to compare the MFB outputs between the reference ($m_{ij}^r(\tau)$) and the test ($m_{ij}^t(\tau)$).
We used the following modified version of cosine similarity:

\begin{eqnarray}
  S_{ij} &=& 
  \frac{\sum_{\tau} w_i(\tau)\cdot m_{ij}^r(\tau)\cdot m_{ij}^t(\tau)} {(\sum_{\tau}  {m_{ij}^r(\tau)}^2)^{\rho} \cdot 
  (\sum_{\tau} {m_{ij}^t(\tau)}^2)^{\,(1-\rho)}}
  \label{eq:GESI_Similarity}
\end{eqnarray}
where $i$ is the GCFB channel, $j$ is the MFB channel, $\tau$ is a frame number. $\rho$ $\{\rho \,|\, 0 \le \rho \le 1\} $ is a weight value that allows us to handle the level asymmetry of the reference and test sounds as explained in section \ref{sec:IntroRho} although 
$\rho=0.5$ in the original definition of cosine similarity.
$w_i(\tau)$ is a weighting function derived from the SSI weight (Size Shape Image weight) explained in section \ref{sec:IntroSSIweight}.
The similarity metric, $d$, is a weighted sum of $S_{ij}$ as follows:
\begin{eqnarray}
   d & = & \frac{1}{MN} \sum_{i=1}^N \sum_{j=1}^M  w_j\, S_{ij},
  \label{eq:GESI_metric}
\end{eqnarray}
where $w_j$ is a weight value that is unity in the current simulation, but is adjustable.

\subsubsection{Introduction of a parameter $\rho$}
\label{sec:IntroRho}

By setting $\rho$ to a value other than 0.5, it is possible to handle the level difference in the MFB outputs, 
$m_{ij}^r(\tau)$ and $m_{ij}^t(\tau)$.
For example, if the reference signal is original clean speech and the test signal is the output of an HL simulator (see section \ref{sec:SubjectiveExp}) or an attenuator, the average levels at $m_{ij}^t(\tau)$ will be smaller than at $m_{ij}^r(\tau)$. This difference should be reflected in a lower SI value. This is not possible when $\rho=0.5$, since the RMS values of the vectors $m_{ij}^r(\tau)$ and $m_{ij}^t(\tau)$ are normalized to unity, resulting in a high similarity and a high SI value. 
It is worth noting that the Pearson correlation used in STOI, ESTOI, and MBSTOI has a similar problem, which will be shown in section \ref{sec:EvaluationOIM}.

In preliminary simulations, we tried several methods to deal with the level difference. For example, we used gain and added a certain level above the absolute threshold, but the resulting values were not stable across the speech data. Introducing $\rho$ was much more stable than any of the other methods we tried.
More importantly, $\rho$ plays a crucial role in reflecting estimated listening level of speech sounds in the SI prediction. The details are described later in section \ref{sec:Predict_Challenge}.

\subsubsection{Introduction of the SSI weight}
\label{sec:IntroSSIweight}

The weighting function $w_i(\tau)$ was determined based on the knowledge from modeling psychoacoustic experimental results on human size perception~\citep{irino2017auditory,matsui2022modelling}.
It was necessary to introduce a weighting function called ``Size Shape Image (SSI) weight'' into the model in order to properly predict the experimental results for male and female speech sounds with different fundamental frequencies, $F_o$.
The SSI weight was developed based on the theory of ``Stabilized Wavelet Mellin Transform (SWMT)'' ~\citep{irino2002segregating} to segregate information about the vocal tract and glottal vibration. 
Recently, it has been shown that introducing the SSI weight into the spectral representation improves the estimation of vocal tract lengths of vowels measured by magnetic resonance imaging (MRI)~\citep{irino2023auditory}.
Therefore, the SSI weight is an effective method for extracting vocal tract information by reducing the effect of the glottal vibration.

The SSI weight was introduced in GESI in the previous study~\citep{irino2022speech} and was a fixed at the average male $F_o$ to predict the SI of male speech experiments. In the current study, we also conducted SI experiments on female speech, in which $F_o$ values are very different from those of male speech. Moreover, $F_o$ varies during the pronunciation.
Therefore, we extended the SSI weight to be a frame-based function, $w^{(SSI)}_i(\tau)$, and used it to determine $w_i(\tau)$ as
\begin{eqnarray}
  w^{(SSI)}_i(\tau) & = & \min(\frac{f_{p,i}}{h_{max}\cdot  F_o(\tau)}, 1), \\ \nonumber
  w_i(\tau) & = & w^{(SSI)}_i(\tau)/ \sum_{i=1}^N w^{(SSI)}_i(\tau),
  \label{eq:SSIweight}
\end{eqnarray}
where $f_{p,i}$ is the peak frequency of the $i$-th GCFB channel and $h_{max}$ is an upper limit parameter. $F_o(\tau)$ is the fundamental frequency of the reference sound at $\tau$. This is estimated by the WORLD speech synthesizer \citep{morise2016world}. If there is no fundamental frequency, as with some consonants, $F_o$ is set to a small positive value close to zero to make $w_i^{(SSI)}$ unity for all $i$.

The above algorithm incorporates unique features of $\rho$ and the SSI weight to improve the SI prediction as shown in section ~\ref{sec:EvaluationOIM}.

\subsection{Evaluation with conventional models}

We have compared GESI with the conventional OIMs: STOI, ESTOI, MBSTOI, HASPIv1, and HASPIv2 in the evaluation (section ~\ref{sec:EvaluationOIM}).
For the evaluation, it is essential to provide a conversion function from the metric to the SI to explain the subjective SI data described in section~\ref{sec:SubjectiveExp}.
We explain the parameters of the conventional OIMs and the conversion function.

\subsubsection{STOI, ESTOI, and MBSTOI}
\label{sec:Eval_Obj_STOI}

STOI \citep{taal2011algorithm} is one of the most popular metrics for evaluating SE algorithms.
The initial STOI process involves one-third octave band analysis, envelope extraction, and calculation of the short-time correlation between the envelopes of the reference and test sounds in each octave.
Then, the internal metric, $d$, is obtained by averaging the inner products between subband temporal envelopes. ESTOI \citep{jensen2016algorithm} shares its envelope extraction with STOI. The metric, $d$, is instead calculated from the average of the correlation coefficients between short-time spectra across subbands. MBSTOI ~\citep{andersen2018refinement} is a binaural extension of ESTOI.

\subsubsection{Conversion from the metric to the SI}
\label{sec:metric to SI}

In STOI, ESTOI, and MBSTOI, 
the metric value, $d$, was converted into word correct rate (\%) or intelligibility, $I$, by a sigmoid function. 
The conversion is performed by a sigmoid function:
\begin{eqnarray}
   I & = & \frac{100}{1+\exp(a\cdot d + b)}.
   \label{eq:STOIGESI_sigmoid}
\end{eqnarray}   
where $a$ and $b$ are parameters (see Eq.\,8 of \citep{taal2011algorithm}, Eq.\,10 of \citep{jensen2016algorithm}, and Eq.\,17 of \citep{andersen2018refinement}).
The same function can be used for GESI, since
GESI also provides a single metric $d$ in Eq.~\ref{eq:GESI_metric}.
The parameter values of $a$ and $b$ are determined from a subset of the SI values in the experimental results using the least squares error (LSE) method.

Note that this conversion is not necessarily required to evaluate SE algorithms. For this purpose
the metric $d$ can be used directly, since the conversion function of Eq. \ref{eq:STOIGESI_sigmoid} is a simple monotonic sigmoid function.

\subsubsection{HASPIv1}
\label{sec:Eval_Obj_HASPIv1}

HASPIv1 was designed to predict SI for HI listeners using hearing aids ~\citep{kates2014hearing}.
HASPIv1 uses an extended version of the gammatone filterbank
and computes two types of features: the cepstral correlations ($c$) and the three levels of auditory coherence ($a_{low}$, $a_{mid}$, and $a_{high}$). 
The SI value is derived using a logistic function, 
\begin{eqnarray}
    I &  = & \frac{100}{1+\exp(-p)}, 
    \label{eq:HASPI_sigmoid} \\
    p & = &\!B\!+\!C\!\cdot c\!+  \!A_{low}\!\cdot\!a_{low}\!+  \!A_{mid}\!\cdot\!a_{mid}\!+ \!A_{high}\!\cdot\!a_{high}. \label{eq:p_HASPI}
    \label{eq:HASPI_linearAdd}
\end{eqnarray}
where $B$ is a bias value; $C$ and $A$ are coefficients corresponding to the features; the coefficients $A_{low}$ and $A_{mid}$ are usually set to zero as in Eqs.\,1 and 7 in \cite{kates2014hearing}. 
The remaining coefficients (i.e., $B$, $C$, and $A_{high}$) are determined from a subset of the SI values in the experimental results by using the LSE method.

The sigmoid function in Eq.~\ref{eq:HASPI_linearAdd} and parameter estimation are essential when using HASPIv1 to evaluate SE algorithms. This is because there are several parameters (i.e., $c$ and $a$) that can act independently and cannot be used as a simple monotonic value that is highly correlated with the SI.

\subsubsection{HASPIv2}
\label{sec:Eval_Obj_HASPIv2}

HASPIv2 is an extended version of HASPIv1 to improve the SI prediction performance~\citep{kates2021hearing}. HASPIv2 provides two types of output: a single SI value as the output of an NN trained using the sentence databases of HINT~\citep{nilsson1994development} and IEEE ~\citep{rothauser1969ieee}, and ten raw parameters, some of which are the same as in HASPIv1, derived from signal analysis.

The NN output is a single value and therefore could be used to evaluate SE algorithms by itself.
It is necessary to convert the NN output to predict the SI of words in the current experiment. This could be done by a sigmoid function in Eq.~\ref{eq:STOIGESI_sigmoid} if they are highly correlated. This method was already used in SI prediction of HI listeners for various situations in the second Clarity Prediction Challenge (CPC2)~\citep{claritychallenge}.

The appropriate method for the current evaluation, which is to organize ten raw parameters into a single metric, was not provided in \cite{kates2021hearing}. 
It may be necessary to train the mapping function with a relatively large word database, but that is beyond the scope of this paper.
Here we have evaluated the performance of HASPIv2 using the NN output, as this is the simplest way for immediate use in evaluating SE algorithms.


\section{Subjective SI experiments}
\label{sec:SubjectiveExp}
Subjective SI experiments were conducted to evaluate the prediction of GESI compared to other methods.
These experiments were listening tests of speech in multi-talker babble noise. They were conducted in the laboratory and in crowdsourced remote environments. In this section, we first describe 
the experimental design and its motivation.
We then describe the speech materials used in the experiments and the process of HL simulation on these sounds. We then describe the laboratory and remote experimental procedures and conditions, followed by the human results.

\subsection{Design of experiments for evaluation}
\label{sec:DesignExp}

The design of SI experiments is a very important issue in the evaluation of OIMs for SE algorithms. Three issues need to be considered: the effect of cognitive factors, individual differences between HI listeners, and individual listening conditions.

\subsubsection{Effect of cognitive factors}
\label{sec:EffectCognitive}

There are several linguistic levels to be tested: phoneme, syllable, word, and sentence. Sentence SI can be strongly influenced by cognitive factors because the sentence consists of a set of words with varying degrees of familiarity: there may be commonly used words and difficult keywords
However, current SE algorithms based on signal processing techniques do not necessarily aim directly at improving cognitive understanding.
Therefore, we conducted SI experiments with words whose familiarity was well controlled to minimize the effect of lexical or cognitive factors ~\citep{sakamoto2006new,kondo2011spoken}.

\subsubsection{Individual difference of HI listeners}
\label{sec:IndivDiffHI}

The degree of hearing loss (HL) and the cognitive factor vary widely among individual HI listeners. Even if we could collect SI data from a large number of HI listeners, it may be difficult to use the SI results (perhaps with clustering) for evaluation because there is no complete test to specify the cognitive factors.
Our approach was to perform SI experiments using an HL simulator for NH listeners whose cognitive function could be assumed to be normal and whose variability could be assumed to be less than that of HI listeners.

For this purpose, we used the latest version of the Wakayama-University Hearing Impairment Simulator, WHIS\,(\citealp{irino2013accurate, nagae2014hearing, irino2020gammachirp, irino2023hearing}), which synthesizes sounds of reasonably high quality that can be used for speech perception experiments \citep{matsui2016effect, irino2020speech,irino2022speech}.
There is a long history of HL simulator development and one of the most popular is CamHLS (\citealp{baer1993effects, nejime1997simulation, stone1999tolerable}).
WHIS was shown to outperform CamHLS in terms of the spectral distance between the auditory model outputs of the HI and the NH with the HL simulator \citep{irino2023hearing}.


\subsubsection{Individual listening conditions}
\label{sec:IndivListenCond}

SI is known to be strongly influenced by the listening environment. For example, SI is usually different when a listener is in a library versus a train station, where the ambient noise levels are different. The quality of audio device can also affect SI.
To evaluate the robustness of OIMs, we conducted SI experiments in a well-controlled laboratory environment and in crowdsourced remote environments. There is a large variability in crowdsourced remote SI experiments~\citep{cooke2011crowdsourcing, paglialonga2020automated,padilla2021binaural,yamamoto2021comparison, irino2022speech} because it is almost impossible to control the listening level, ambient noise level, individual audiogram, and equipment even when NH listeners participate. The objective SI prediction of individual NH listeners in these situations would serve as a good initial test for that of individual HI listeners.

\subsection{Speech materials}
\label{sec:SpeechMatrial}

\subsubsection{Speech data}
\label{sec:SourceSpeech}

The speech sounds used for the subjective listening experiments were Japanese 4-mora words. The Japanese mora is a unit of speech that roughly corresponds to the CV syllable, except for a few special ones. They were uttered by a female speaker ``fhi'' and a male speaker ``mis'', drawn from the lowest familiarity set in a database of familiarity-controlled word lists, FW07 \citep{sakamoto2006new,kondo2011spoken}.
The dataset contained 400 words for each of the four familiarity ranks, and the average duration of a 4-mora word was approximately 700\,ms.
Using the lowest familiarity rank avoids inflating the SI value by guessing frequently used words. This is an effective way to reduce the effect of individual differences in mental lexicon.

\subsubsection{Noise conditions}
\label{sec:SourceNoise}
To perform speech-in-noise tests,
babble noise was added to the clean speech to create noisy speech sounds. This is referred to as ``unprocessed'' because there were other post-processed conditions described later.
The SNR conditions ranged from $-3$\,dB to $+9$\,dB in $3$-dB steps.

Babble noise has a temporal fluctuation in power that reduces the intelligibility of individual speech. This babble noise was generated from the word sounds contained in the FW03 database ~\citep{amano2009development} as follows: the word sounds were randomly selected and concatenated to be 5 minutes long; it was repeated 32 times with different word; 32 sets of sound data were added together with random starting time to produce a single track sound. 32 was chosen so that the verbal information of each speech would not be discernible, and so that the mixed sound would not be a steady noise like white noise. 
For noisy sounds, babble noise of the same length as the original speech sound was extracted from this long babble noise from a random starting point. 
All sounds were processed at a sampling rate of 48\,kHz.

\begin{table}[t]
\vspace{10pt}
 \caption{Average hearing levels (dB) of 70-year-old male listeners~\citep{iso7029} and 80-year-old male and female listeners \citep{tsuiki2002nihon}. The value for 80-year-old and 6000\,Hz in a parenthesis is an interpolated value used in HASPIv1. }\label{tab:HL}
 \centering
  \begin{tabular}{|c||c|c|c|c|c|c|c|c|c|}
   \hline
    Freq.& 125 & 250 & 500 & 1000 & 2000 & 4000 & 6000 & 8000\\
   \hline \hline 
    70-yr & 8 & 8 & 9 & 10 & 19 & 43 & 49 & 59 \\
    \hline 
   80-yr & 24 & 24 & 27 & 28 & 33 & 48 & (58) & 69 \\
    \hline
  \end{tabular}
\end{table}

 
\subsubsection{HL simulation}
\label{sec:HearingLossSimulation}

The noisy sounds described above were further processed to derive simulated HL sounds using WHIS, which is described in 
\citep{irino2023hearing}.
Briefly, WHIS first analyzes input sounds with the gammachirp auditory filterbank (GCFB), which computes excitation patterns (EPs) based on the hearing level that appears in the audiogram and the compression health, which is closely related to loudness recruitment~\citep{moore2013introduction} of an HI listener. Then, WHIS synthesizes simulated HL sounds from the difference between the EPs of HI and NH listeners using a direct time-varying filter (DTVF) method that does not produce large distortions. 


Table~\ref{tab:HL} shows the average hearing levels used for the simulation: 70-year-old male listeners (hereafter ``70-yr''), as defined in \cite{iso7029}, and 80-year-old listeners (hereafter ``80-yr''), as defined in \cite{tsuiki2002nihon}.
We set the compression health to 0.5 to simulate moderate dysfunction in the active process in the cochlea. 

In addition, we added a condition in which the sound pressure level (SPL) of the source sounds was simply reduced by 20\,dB (hereafter ``low-level'') to clarify the difference in SI between flat level reduction and high frequency HL. 
This reduction level was chosen because the simulated 80-yr sounds were approximately 20\,dB lower than the source sounds. 


\subsection{Experimental procedure}
\label{sec:SbjSI_Procedure}
We had developed a set of web pages that could be used in both laboratory and remote SI experiments in \cite{yamamoto2021comparison}. Google Chrome was chosen as the usable browser because it plays 48-kHz and 16-bit wav files correctly on Windows and Mac systems. Participants were required to read information about the experiments before giving informed consent by clicking the consent button twice to ensure their agreement. The experiments were approved by the ethics committee of Wakayama University (Nos. 2015-3, Rei01-01-4J, and Rei02-02-1J).
They then entered the questionnaire page, which included questions about age, type of wired headphones or wired earphones (use of Bluetooth or a loudspeaker was not permitted) and native language (Japanese or not), as well as self-report of HL (yes or no). The conditions for audio equipment are described in sections \ref{sec:SbjSI_Lab} and \ref{sec:SbjSI_Remote}.
They then took the tone pip tests, as described in section \ref{sec:TonePip}~\citep{yamamoto2022effective}.
Next, the participants completed a training session in which they performed an easy task using the same procedure as in the main experimental sessions to familiarize themselves with the tasks. Here, the speech sounds were drawn from words with the highest familiarity rank and with an SNR above 0\,dB.
Participants were instructed to write down the words they heard using hiragana (i.e.~Japanese characters) during four-second pauses between words.
Then the main experiment began.
The total number of stimuli presented was 400 words, consisting of a combination of four HL conditions 
\{unprocessed, 70-yr, 80-yr, and low-level\} and five SNR conditions with 20 words per condition. There were 40 sessions of 10 words each. Each participant listened to a different set of words, which were randomly assigned to avoid bias due to word difficulty. 
The experiment was divided into two one-hour tasks to meet the crowdsourcing requirement of the task duration.


\subsubsection{Leading sentence for familiarization with the sound level}
\label{sec:LeadingSentence}
In each session, we introduced the following leading sentence: ``Speech sounds will be presented at this volume'' in Japanese. This was followed by 10 test words. The sentence and the words were processed in the same HL condition. In the preliminary experiments, when the words were randomly presented at different sound levels, it was not easy for listeners to concentrate on the sounds because they were trying to avoid being startled by a louder sound immediately after a softer sound. In fact, the leading sentence helped the listeners to concentrate.


\subsubsection{Tone pip test for estimating listening conditions}
\label{sec:TonePip}

In crowdsourced remote experiments, it is impossible to strictly control the listening environments. For example, participants could listen to the stimulus sounds in different environments even though they were instructed to perform the task in a ``quiet'' place. In addition, it is difficult to obtain the information about the sound presentation level, the ambient noise level, the quality of the audio equipment, and the hearing level. However, it is known that the sensation level in the given listening environment affects the SI~\citep{french1947factors}.

\begin{figure}[t]
    \centering
    \includegraphics[width = 1\linewidth] {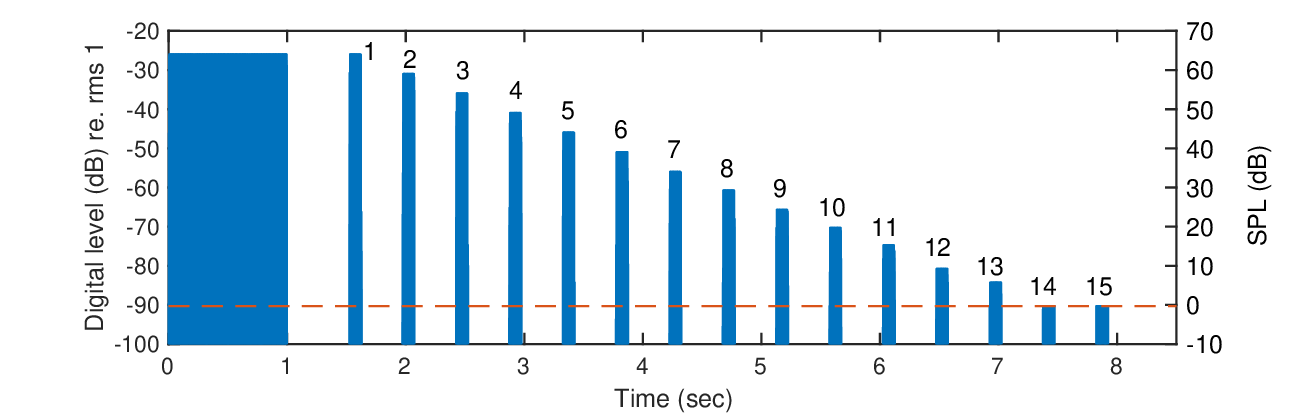}
    \caption{
    RMS digital level of a sequence of 15 tone pips decreasing in steps of 5\,dB. The right y-axis shows the SPL when the first pip is assumed to be 64\.dB SPL.
    }
    \label{fig:TonePip}
\end{figure}

For this analysis, we also introduced tone pip tests to estimate how much the speech sounds were presented above the just audible level.
A sequence of 15 tone pips with decreasing steps of -5\,dB was presented to listeners, who were asked to report the number of audible tone pips, $N_{pip}$. This is a simplified version of a sensation level measurement in psychoacoustic experiments ~\citep{moore2013introduction}.
Figure~\ref{fig:TonePip} shows the RMS digital level of the sequence of tone pips following a 1-second reference tone sound that has the same SPL as the stimulus speech sounds. 
The tone frequencies were 500, 1000, 2000, and 4000\,Hz to cover the speech range.
The SPLs at hearing level of 0\,dB are 13.5, 7.5, 9.0, 12.0\,dB at these frequencies~\citep{ANSI_S3.6-2010}. The mean value is 10.5\,dB and the standard deviation is 2.7\,dB which is smaller than the step size of 5\,dB. Therefore, we used the average value over the frequencies, $\bar N_{pip}$, to analyze the relationship with the SI. 
This $\bar N_{pip}$ value provides a rough indication of the sound level above the audible threshold or margin in the acoustic environment of the individual participant as
\begin{eqnarray}
     L_{sm} & = & 5 \cdot(\bar N_{pip}-1).
     \label{eq:Llis=5Npip}
\end{eqnarray}
The right y-axis in Fig.~\ref{fig:TonePip} shows an example when the stimulus SPL is 64\,dB.

One of the most important findings of this study is that $\bar N_{pip}$ can be reflected to determine the parameter $\rho$ in Eq.~\ref{eq:GESI_Similarity} as described later in section \ref{sec:Predict_Challenge}.
The tone pip test took only a few minutes. The procedure is very simple and improved by using an ascending sequence together. It was also possible to use $\bar N_{pip}$ for data screening effectively~\citep{yamamoto2022effective}.

\subsection{Laboratory experiments}
\label{sec:SbjSI_Lab}
The laboratory SI experiment for male speech sounds was conducted first, with thirteen young NH listeners (aged 20--23 years). The laboratory SI experiment for female speech sounds was conducted after the remote experiments for male speech (section \ref{sec:SbjSI_Remote}), in which fourteen NH listeners (aged 19--23 years) participated. 
%
%
%
The listeners were seated in a sound-attenuated room with a background noise level of approximately 26dB in $L_{\rm Aeq}$. They were all Japanese and naive to our SI experiments and had a hearing level of less than 20\,dB between 125\,Hz and 8,000\,Hz.
The sounds were presented diotically through a DA-converter (SONY, NW-A55) via headphones (SONY, MDR-1AM2). 
The SPL of the unprocessed sounds was 65\,dB in ${L_{eq}}$, which was the same level as the calibration tone measured with an artificial ear (Br\"{u}el \& Kj\ae r, Type\,4153), a microphone (Br\"{u}el \& Kj\ae r, Type\,4192), and a sound level meter (Br\"{u}el \& Kj\ae r, Type\,2250-L).

\subsection{Remote experiments}
\label{sec:SbjSI_Remote}

The same SI experiments for male and female speech sounds were outsourced to a crowdsourcing service provided by Lancers Co. Ltd. in Japan \citep{Lancers} as in \cite{yamamoto2021comparison} after the corresponding laboratory experiments. 
In the male speech experiment,
any crowdworker could participate in the experimental task on a first-come, first-served basis. Data from twenty-seven listeners (aged 22--66 years old) were used as the experimental results after removing incomplete responses.
In the female speech experiment, a pre-screening experiment described in appendix \ref{secApndx:Pre-screening} was conducted in advance on a first-come, first-served basis.
We then asked 95 pre-screened crowdworkers to participate in the experiments and obtained data from twenty-nine participants (aged 23--57years old). 
Participants in both experiments were all Japanese and naive to our SI experiments. None of them self-reported any hearing impairment.

Participants were asked to perform the experiments in a quiet place and to set the volume of their headphones or earphones to a comfortable level for the unprocessed condition and to a tolerably audible level for the low-level condition.
It was difficult to control the listening conditions more precisely. However, some of the conditions could be roughly estimated using the tone pip tests described in section \ref{sec:TonePip}. 

\begin{figure*}[t]
\parskip=0pt
\baselineskip=0pt
\figline{
\fig{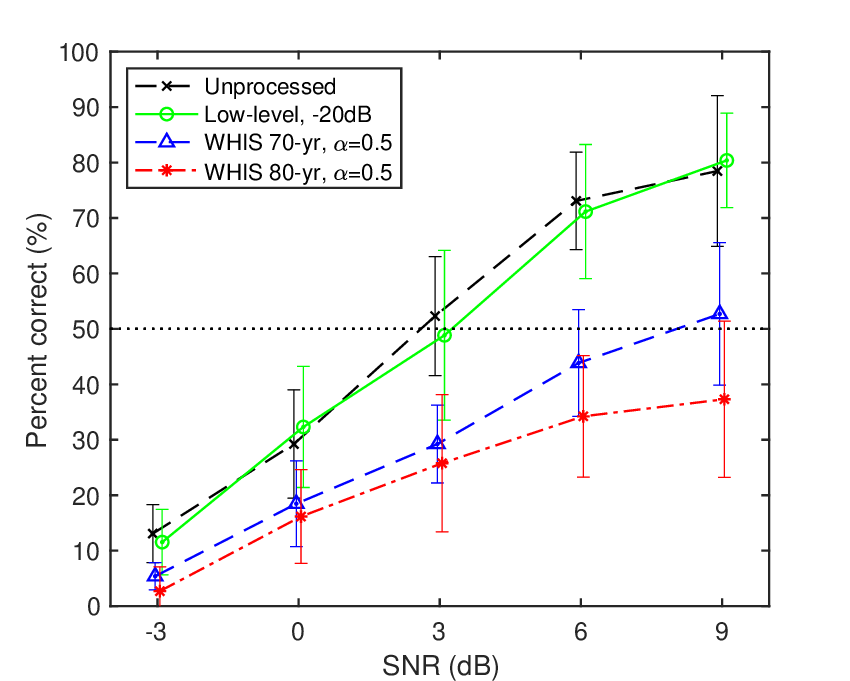}{0.45\linewidth}{\label{fig:RsltSbj_mis_Lab} (a) Human: Male, Laboratory}
\fig{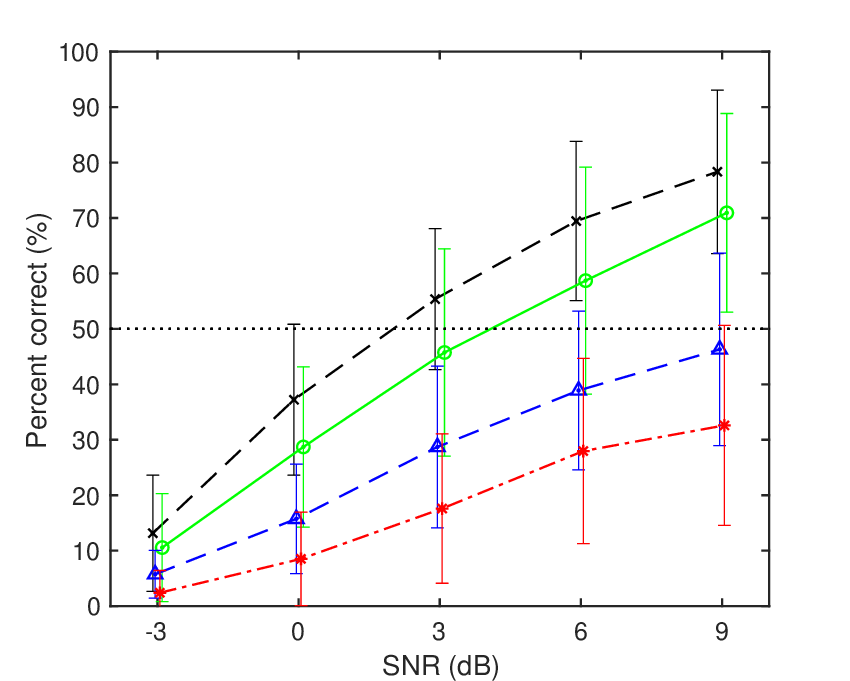}{0.45\linewidth}{ \label{fig:RsltSbj_mis_Remote} (b) Human: Male, Remote}
}
\vspace{-12pt}

\figline{
\fig{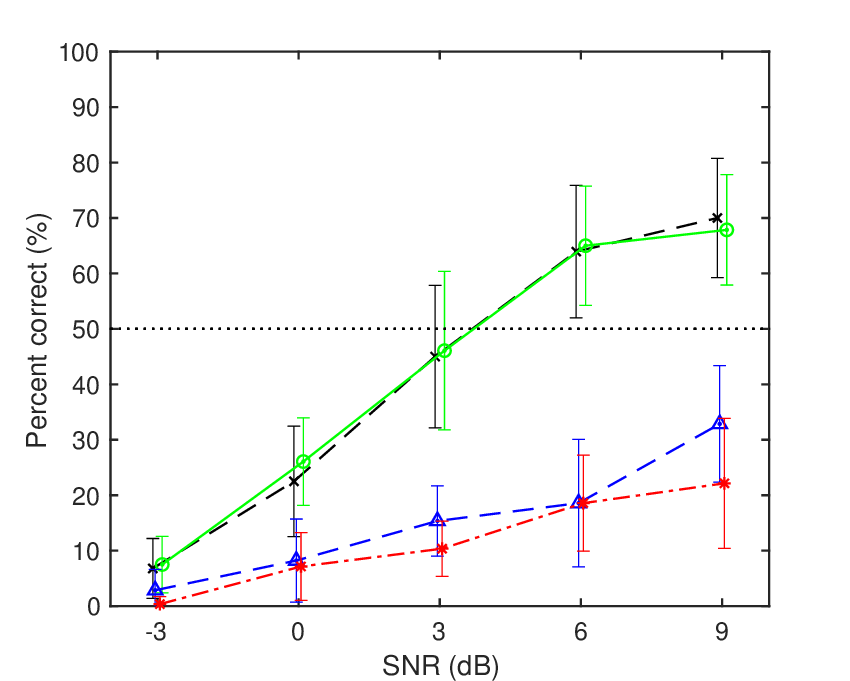}{0.45\linewidth}{\label{fig:RsltSbj_fhi_Lab} (c) Human: Female, Laboratory}
\fig{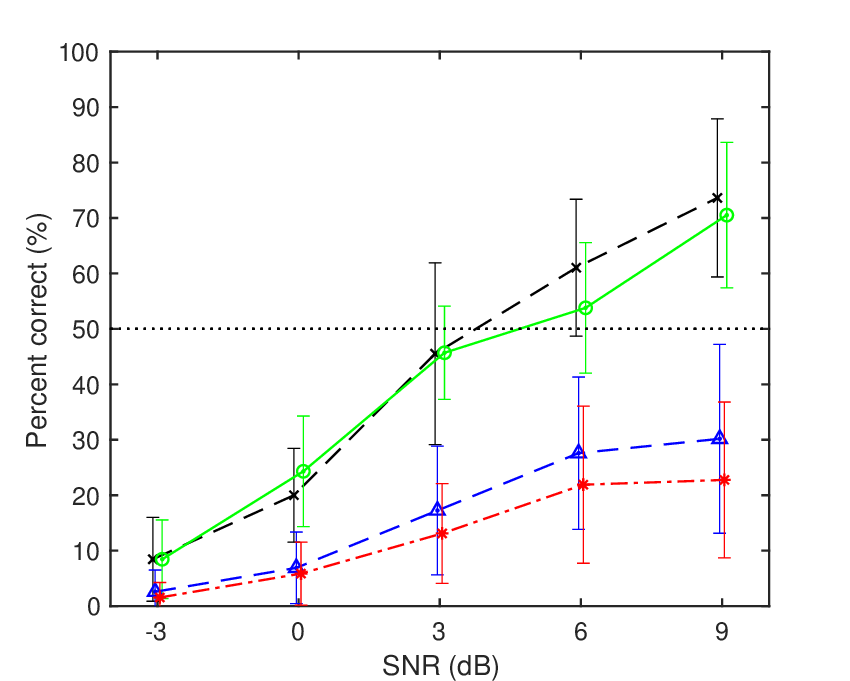}{0.45\linewidth}{\label{fig:RsltSbj_fhi_Remote} (d) Human: Female, Remote}
}

\caption{\label{fig:RsltSbj} Subjective SI results: Mean and standard deviation (SD) of word correct rate (\%) across listeners.
    }
\end{figure*}

\subsection{Experimental results}
\label{sec:SbjSI_Result}

Figure~\ref{fig:RsltSbj} shows the subjective SI values, defined as the word correct rates, as a function of the SNR. Circles and error bars represent the mean and standard deviation (SD) across participants. 

In the laboratory experiments on male speech (Fig.~\ref{fig:RsltSbj}(a)), the lines of the unprocessed and low-level conditions were almost the same. Therefore, the level reduction of 20\,dB did not affect SI in the well-controlled experiments with the young NH listeners. However, in the remote experiments (Fig.~\ref{fig:RsltSbj}(b)), the line of the low-level condition was lower than that of the unprocessed condition, indicating that the 20\,dB gain reduction affected the SI. 

In the laboratory experiments of female speech(Fig.~\ref{fig:RsltSbj}(c)), the difference between the unprocessed and low-level conditions was also almost the same. In the remote experiment (Fig.~\ref{fig:RsltSbj}(d)), the difference was smaller than that observed in the male remote experiment. This is probably due to the effect of the pre-screening conducted before the remote experiment.


The SI values in the 70-yr and 80-yr conditions were higher in the male experiments than in the female experiments, as expected. However, it is also the case that the SI values of the unprocessed and low-level conditions were slightly higher for the male speech than for the female speech, likely due to the difference in the listenability of the original speech. 
Overall, the SDs were larger in the remote experiments than in the laboratory experiments, likely due to the different listening conditions of the individual participants.

The above observation was not statistically tested because the individual difference and its cause are more important than the population argument, as described in the next section. Also, the individual SI values, not the average, were used to evaluate the OIMs in section \ref{sec:EvaluationOIM}.


\begin{figure*}[t]
\parskip=0pt
\baselineskip=0pt
\figline{
\fig{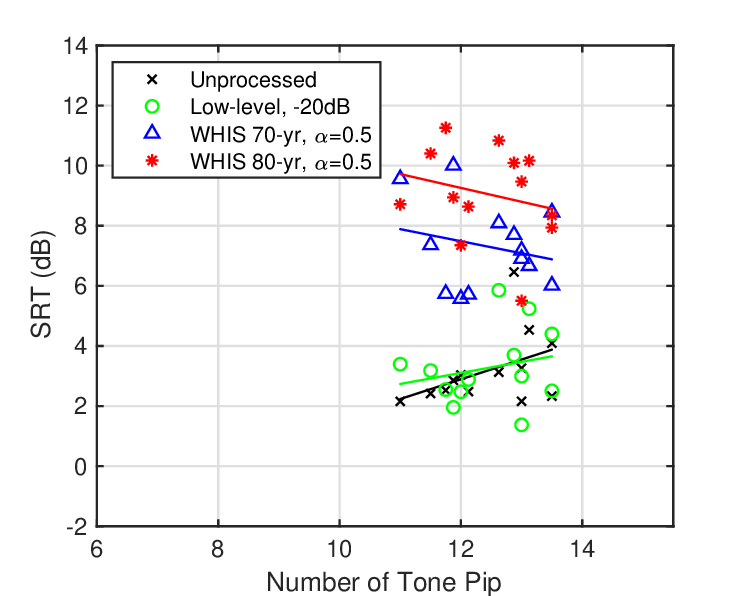}{0.45\linewidth}{\label{fig:Scatter_NpipSRT_male_Lab} (a) Male, Laboratory }
\fig{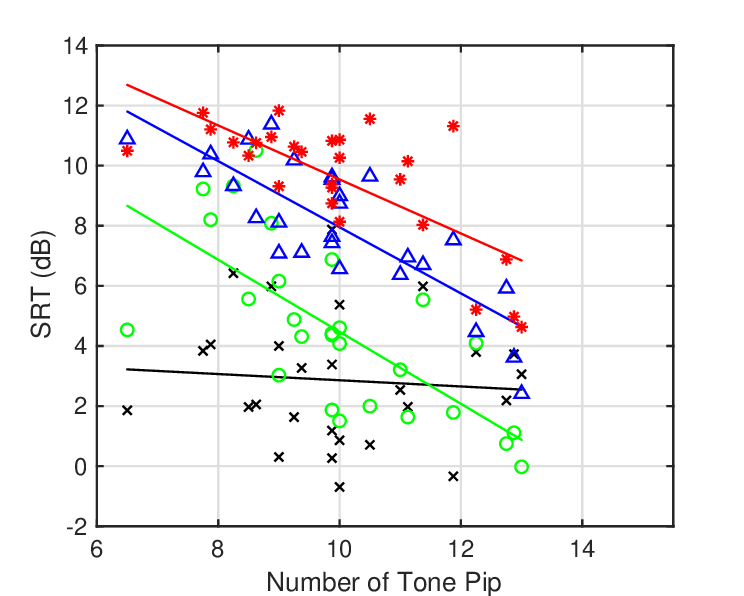} 
{0.45\linewidth}{ \label{fig:Scatter_NpipSRT_male_Remote} (b) Male, Remote}
}

\figline{
\fig{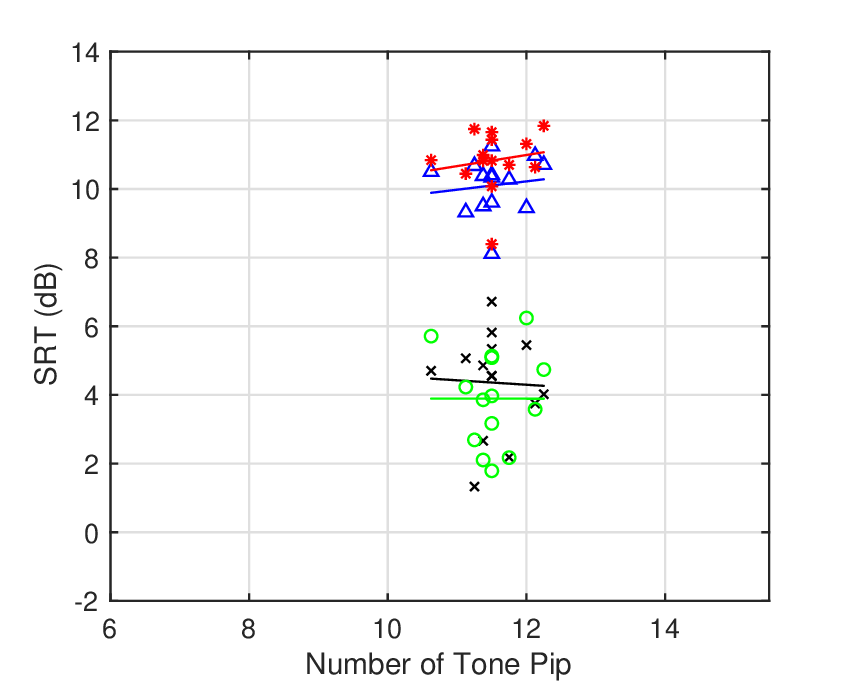}{0.45\linewidth}{\label{fig:Scatter_NpipSRT_female_Lab} (c) Female, Laboratory}
\fig{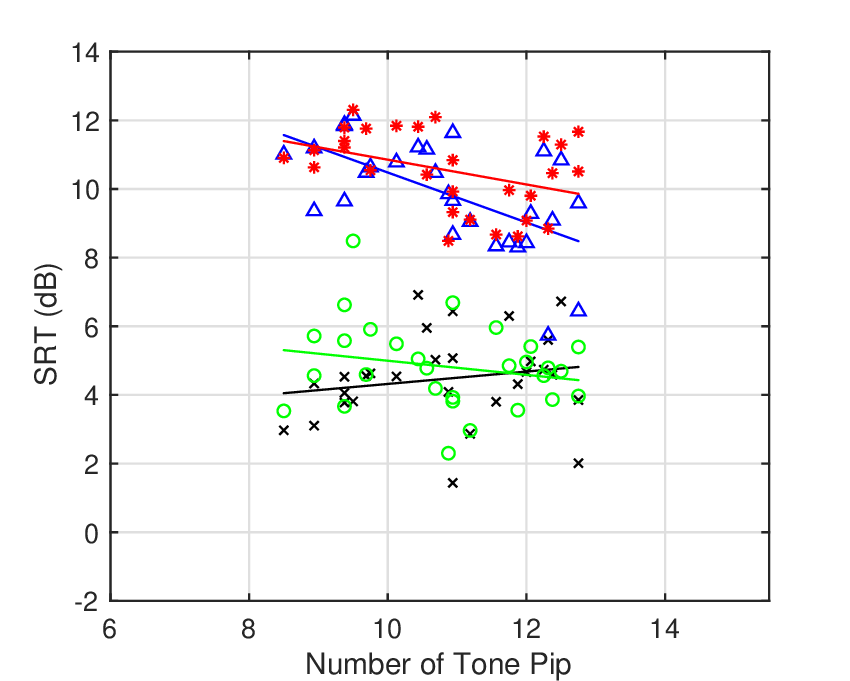}{0.45\linewidth}{\label{fig:Scatter_NpipSRT_female_Lancers} (d) Female, Remote}
}

    \caption{ \label{fig:Scatter_NpipSRT} Scatter plot for the mean reported number of audible tone pips, 
    $\bar N_{pip}$, versus the mean SRT value (dB) for the male and female speech experiments. The conditions were unprocessed (black), low-level (green), 70-yr (blue), and 80-yr (red). The experiments of the panels correspond to those in Fig.~\ref{fig:RsltSbj}.
    Each point represents an individual listener. The solid lines are the regression results.
    }

\end{figure*}

\subsubsection{Relationship between listening condition and SI}
\label{sec:SRTvsTonePiP_WHIS}

The effect of listening environment on intelligibility was investigated in more detail. The speech reception threshold (SRT) was calculated from the SI results in Fig.~\ref{fig:RsltSbj} as the SNR value at which the psychometric function reaches a 50\% word correct rate for each participant and each speech condition.
If there is any relationship between the SRT values and the reported number of tone pips, $\bar N_{pip}$ (section \ref{sec:TonePip}), it is possible that the listening condition is affecting the SI value.
Figure \ref{fig:Scatter_NpipSRT} shows a scatter plot between the reported number of the tone pip averaged over four tone frequencies ($\bar N_{pip}$) and the mean SRT value (dB) averaged over the four HL conditions. Each point represents an individual listener. 

\paragraph{Results for male speech}
\label{sec:SRTvsTonePiP_WHIS_male}

In the laboratory experiment of male speech (Fig.~\ref{fig:Scatter_NpipSRT}(a)), $\bar N_{pip}$ was ranged from 10 to 14 and was not significantly correlated with the mean SRT value.
On the other hand, the remote experiment (Fig.~\ref{fig:Scatter_NpipSRT}(b)) yielded a different result than the laboratory experiment. There was no significant correlation in the unprocessed condition (black; $r=-0.079; p=0.70$). However, there were highly significant correlations in the low-level (green; $r=-0.70; p \ll 0.001$), 70-yr (blue; $r=-0.82; p \ll 0.001$), and 80-yr (red; $r=-0.73; p \ll 0.001$) conditions.


%
When $\bar N_{pip}$ was less than 9 as in Fig.~\ref{fig:Scatter_NpipSRT}(b), 
the dynamic range above the threshold was less than 40\,dB ($= 5 \times (9-1)$ from Eq.~\ref{eq:Llis=5Npip}). This margin did not seem sufficient to detect low level consonants even in the low-level condition, which is a flat level reduction. This is probably one reason why the difference between the unprocessed and low-level conditions was much larger in the remote experiment (Fig.~\ref{fig:Scatter_NpipSRT}(b)) than in the laboratory experiment (Fig.~\ref{fig:Scatter_NpipSRT}(a)).
$\bar N_{pip}$ of less than 9 is also observed for the 80-yr and 70-yr conditions in Fig.~\ref{fig:Scatter_NpipSRT}(b), and this may be closely related to the low SI values in Fig.~\ref{fig:RsltSbj}(b).

\paragraph{Results for female speech}
\label{sec:SRTvsTonePiP_WHIS_female}
The above observation in the male experiments led us to develop the pre-screening experiment, performed before the female speech experiment, to limit the range of $\bar N_{pip}$ and to control the audio device somewhat tightly, as described in appendix \ref{secApndx:Pre-screening}.
In the laboratory experiment on female speech (Fig.~\ref{fig:Scatter_NpipSRT}(c)), the range of $\bar N_{pip}$ was between approximately 10.5 and 12.5, i.e.~narrower than that observed in the experiment of male speech (Fig.~\ref{fig:Scatter_NpipSRT}(a)). There was no significant correlation with the mean SRT value.


In the remote experiments (Fig.~\ref{fig:Scatter_NpipSRT}(d)), the $\bar N_{pip}$ range was between approximately 8.5 and 13, i.e.~narrower than in the male speech experiment (Fig.~\ref{fig:Scatter_NpipSRT}(b)).
%
%
There was no significant correlation in the unprocessed (black; $r=0.18; p=0.35$) and low-level (green; $r=-0.21; p=0.27$) conditions. There were significant correlations in the 70-yr (blue; $r=-0.60; p\ll 0.001$) and 80-yr (red; $r=-0.40; p=0.032$) conditions.

Therefore, fewer significant correlations were observed than those in the male speech experiment. This result may reflect the smaller difference between the SI values of the unprocessed and low-level conditions, observed in the remote female speech experiment in Fig.~\ref{fig:RsltSbj}(d) than that observed in the remote male speech experiment in Fig.~\ref{fig:RsltSbj}(b).
This may imply that the pre-screening before the remote female speech experiments worked effectively.

In summary, the tone pip test can provide good information about the listening conditions known to affect the SI. 



\section{Evaluation of OIMs}
\label{sec:EvaluationOIM}

GESI and other OIMs were evaluated in terms of how well they predicted the subjective SI results in section \ref{sec:SbjSI_Result}.

\subsection{Motivation of evaluations}
\label{sec:Predict_motivation}

We performed three types of prediction assessments. Their motivations are described here.
The intrusive OIMs estimate the SI by comparing the test signals with the reference signals, which in this study were the original clean speech sounds. The test signal and parameters were set according to the type of evaluation.

\begin{description}
    \item \textbf{Eval.1:} \textit{ Prediction of the average SI with using simulated HL sounds (Section \ref{sec:Eval_WithWHIS})}
    
    We investigated whether the OIMs could predict the average SI of NH listeners shown in Fig.~\ref{fig:RsltSbj} when using simulated HL sounds which were presented to the listeners.
    There was a difference in SPL between the reference signal and test signal.
    Most of conventional OIMs, except a few such as HASPI, normalize both of the reference and test sounds to the same RMS level. Thus, in theory, they cannot predict the SI difference between the unprocessed and low-level conditions because the RMS levels of their sounds become identical. This would be the case for the HL conditions.

    \item \textbf{Eval.2:} \textit{ Prediction of the average SI without using simulated HL sounds with limited parameter value settings (Section \ref{sec:Eval_WithoutWHIS})}
    
    The main goal of GESI is to develop an SI measure for HI listeners although we introduced WHIS to conduct the experiments for NH listeners in order to avoid the problems that might arise in experiments with HI listeners, as described in section \ref{sec:DesignExp}. Therefore, it is important to predict the SI values of 70-yr and 80-yr conditions without using simulated HL sounds. 
    We investigated how well GESI and HASPI, which can reflect the audiograms of HI listeners, predict the SI values shown in Fig.~\ref{fig:RsltSbj}. STOI, ESTOI, and MBSTOI were not included in the comparison simply because they do not provide a method of introducing the audiograms.
    In addition, we investigated the generalization property of GESI and HASPI. The parameters for the SI sigmoid function (Eqs.~\ref{eq:STOIGESI_sigmoid} and \ref{eq:HASPI_sigmoid}) were determined only from the unprocessed condition of the laboratory experiment for male speech. 
    The prediction was made for both laboratory and remote experiments for both male and female speech sounds.

    \item \textbf{Eval.3:} \textit{ Prediction of the SI of the individual listeners (Section \ref{sec:Eval_Indiv})}
    
    It is important for a new OIM used in personal devices to predict the SI values for individual listeners in different conditions from the minimum number of the SI values for parameter setting. 
    In the current experiments shown in Fig.~\ref{fig:RsltSbj}, there was non-negligible variability between the NH listeners, although it must be less than that for HI listeners. The objective SI prediction for individual NH listeners would serve as a good initial test for that for individual HI listeners.

\end{description}


\subsection{Challenge in the prediction}
\label{sec:Predict_Challenge}

There is a challenge in the prediction.
As shown in Fig.~\ref{fig:RsltSbj}, the difference between the SI values of the unprocessed 
and low-level conditions was greater in the remote experiments than in the laboratory experiments, particularly in the male speech experiments (Figs.~\ref{fig:RsltSbj}(a) and \ref{fig:RsltSbj}(b)).
However, the set of stimulus speech sounds was identical, meaning that the prediction would be virtually the same if the difference between the laboratory and remote experiments were not taken into account.
The different listening conditions of the participants are probably the main cause of the different SI values.
The tone pip tests could provide some of the information described in section \ref{sec:TonePip} and Fig.~\ref{fig:Scatter_NpipSRT}.
GESI was designed to reflect this information in its prediction.

As shown in Fig.~\ref{fig:Scatter_NpipSRT}(b), the SRT increased as the mean reported number of tone pips $\bar N_{pip}$ decreased. Therefore, the metric $d$ (in Eq.~\ref{eq:GESI_metric}) and the SI (Eq.~\ref{eq:STOIGESI_sigmoid}) must be correlated with $\bar N_{pip}$.
We introduced this relationship by assuming that the parameter $\rho$ in Eq.~\ref{eq:GESI_Similarity} is a linear function of $\bar N_{pip}$, as follows:
\begin{eqnarray}
    \rho & = & 0.50 + 0.02\cdot(15 - \bar N_{pip}).
    \label{eq:rho=05}
\end{eqnarray}
The coefficients were determined to reasonably predict the experimental results of the male speech experiments shown in Figs.~\ref{fig:RsltSbj}(a) and \ref{fig:RsltSbj}(b). 
This equation was valid for all predictions in this paper without changing the coefficients.

It is possible to modify Eq.~\ref{eq:rho=05}, for example by using a linear regression fit,
to improve the performance.
However, we restrict ourselves to using this simple equation in this study, as in the previous study~\citep{irino2022speech}. This is because the significant figures of $\bar N_{pip}$ have one decimal place and excessive tuning of the coefficients would not be appropriate for generalization.

Table~\ref{tab:Npip_rho} shows
the mean reported number of tone pips, $\bar N_{pip}$, averaged across participants and the corresponding $\rho$ values calculated by Eq.~\ref{eq:rho=05} for the prediction in Eval.1 (section \ref{sec:Eval_WithWHIS}) and Eval.2 (section \ref{sec:Eval_WithoutWHIS}). For the prediction in Eval.3 (section \ref{sec:Eval_Indiv}), the $\bar N_{pip}$ values were individually different. 



\begin{table}[t]
    \caption{Mean and standard deviation (SD) across listeners of the mean reported number of audible tone pips, $\bar N_{pip}$, and the corresponding $\rho$ value used for prediction in Eval.1 (section \ref{sec:Eval_WithWHIS}). 
    $\rho$ was calculated using Eq.~\ref{eq:rho=05}. 
    }\label{tab:Npip_rho}
    \centering
    \begin{tabular}{lcc}	
    \toprule
      & $\bar N_{pip}$ & $\rho$\\
    \midrule
    Male speech, Lab. &	12.6 (0.8) & 0.55 (0.02) \\
    Male speech, Remote &	 10.0 (1.7) & 0.60 (0.03) \\
    Female speech, Lab. & 11.5 (0.4) & 0.57 (0.01) \\
    Female speech, Remote &	 10.8 (1.3) & 0.58 (0.03) \\
    \bottomrule
    \end{tabular}
\end{table}
\begin{table}[t]
    \caption{Data for setting the parameters of the SI sigmoid functions (Eqs.~\ref{eq:STOIGESI_sigmoid} and \ref{eq:HASPI_sigmoid}) and the derived values used for prediction in Eval.1 and Eval.2.
    }
    \label{tab:Eval_ParamSet_Eval12}
    \centering
    \begin{tabular}{ll}	
    \toprule
        & Data for setting and the values for prediction \\
    \midrule
        Subjective SI & Mean SI across listeners: 5 SNRs \\
        \cline{1-2}
        GESI/STOI & Mean metric across listeners \& words: 5 SNRs \\
        & GESI: $a = -12.20$, $b=6.28$ \\
        & \hspace{18pt} $\rho$ (Eval.1): Mean values listed in Table~\ref{tab:Npip_rho}\\   
        & \hspace{18pt} $\rho$ (Eval.2): Individually different\\
        & STOI: $a = -11.01$ , $b = 8.49$\\
        & ESTOI: $a = -7.59$ , $b = 4.46$\\
        & MBSTOI: $a = -8.76$ , $b = 6.29$\\
        & HASPIv2: $a = -5.91$ , $b = 3.49$ \\ 
        \cline{1-2}
        HASPIv1 & Metric: 13 listeners $\times$ 20 words $\times$ 5 SNRs   \\            
        & $B=-13.81$, $C=1.49$, $A_{high}=17.10$\\
    \bottomrule
    \end{tabular}
\end{table}
\begin{table}[t]
    \caption{Data for setting the parameters of the SI sigmoid functions (Eqs.~\ref{eq:STOIGESI_sigmoid} and \ref{eq:HASPI_sigmoid}) and the derived values used for prediction in Eval.3.}
    \label{tab:Eval_ParamSet_Eval3}
    \centering
    \begin{tabular}{ll}	
    \toprule
        & Data for setting and the values for prediction \\
    \midrule
        Subjective SI & Individual SI: 5 SNRs\\
        \cline{1-2}
        GESI & Mean metric across words: 5 SNRs \\
        & $a$, $b$, $\rho$: Individually different \\
        \cline{1-2}
        HASPIv1 & Metric: 20 words $\times$ 5 SNRs \\
        & $B$, $C$, $A_{high}$: Individually different \\

    \bottomrule
    \end{tabular}
\end{table}


\subsection{\textbf{Eval.1}: Prediction of the average SI with using simulated HL sounds}
\label{sec:Eval_WithWHIS}

We evaluated GESI, STOI, ESTOI, MBSTOI, HASPIv1, and HASPIv2 in terms of predicting the average SI of the male speech experiments shown in Figs.~\ref{fig:RsltSbj}(a) and \ref{fig:RsltSbj}(b).

\paragraph{Parameter settings} The parameters for the SI sigmoid function (Eqs.~\ref{eq:STOIGESI_sigmoid} and \ref{eq:HASPI_sigmoid}) were determined by least squares for the subjective and predicted SI values in the unprocessed condition of the male speech laboratory experiment only. 
Tables~\ref{tab:Eval_ParamSet_Eval12} and \ref{tab:Eval_ParamSet_Eval3} show the summary of the data used to determine the parameters and the derived values used for prediction. The parameter setting and prediction were made using exactly the same words that each participant heard.
The subjective SI values were the mean SI values across listeners, i.e., one value at each SNR and 5 values in total. The objective metrics used in GESI, STOI, ESTOI, MBSTOI, 
and HASPIv2 were the mean values across listeners and words, i.e., also one value at each SNR and 5 values in total. 
In contrast, the objective metrics used in HASPIv1 were values from all 13 listeners and 20 words at each SNR. This is because, unlike GESI and STOI, it was difficult to accurately estimate the parameters $B$, $C$, and $A_{high}$ from only 5 values. Therefore, HASPIv1 requires more variability in the data for parameter setting due to the different properties of cepstral correlations $c$ and auditory coherence $a_{high}$ in Eq.~\ref{eq:p_HASPI}.
The $\rho$ values for GESI used for prediction are listed in Table~\ref{tab:Npip_rho}. 


\begin{figure*}[t]

    \parskip=0pt
    \baselineskip=0pt

\figline{
\fig{Fig3_ExpRsltPC_WHIS_Lab.eps}{0.33\linewidth}{\vspace{-10pt}\label{fig:RsltSbj_mis_Lab_Eval1} (a) Human: Male, Lab.}
\fig{Fig3_ExpRsltPC_WHIS_Remote.eps}{0.33\linewidth}{ \vspace{-10pt}\label{fig:RsltSbj_mis_Remote_Eval1} (b) Human: Male, Remote}

\fig{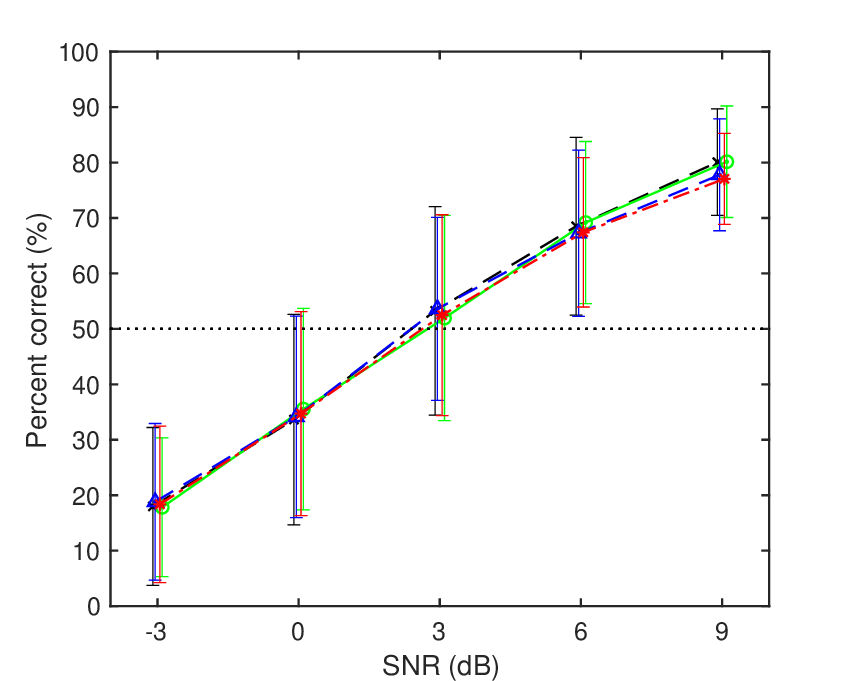}{0.33\linewidth}{\vspace{-10pt}\label{fig:STOI_mis_Lab_OptmMean} (c) STOI 
}
}
\vspace{-12pt}
    \figline{
    \fig{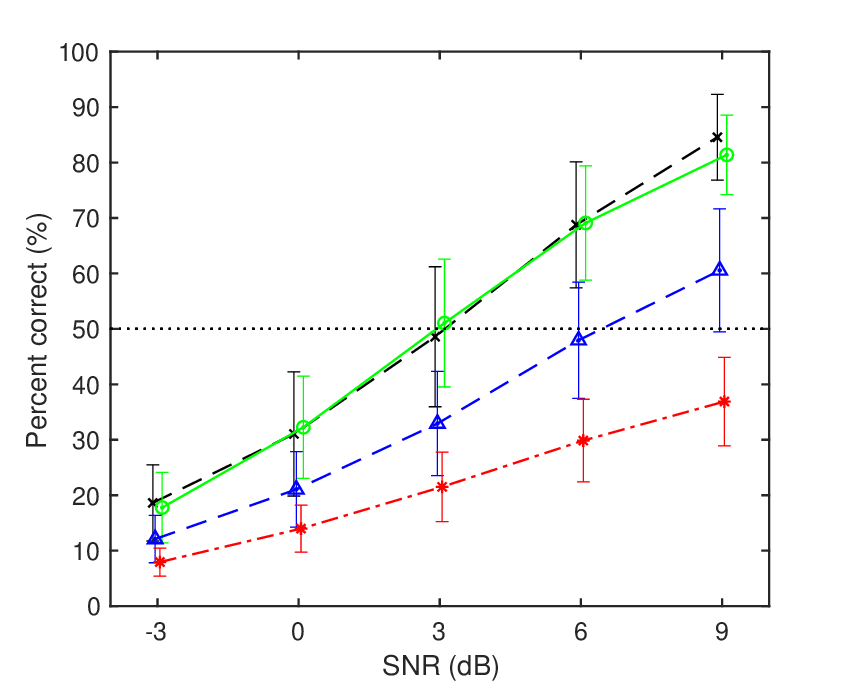}{0.33\linewidth}
    {\vspace{-10pt}\label{fig:GESI_mis_Lab_OptmMean_rPwr55} (d) GESI ($\rho = 0.55$)}
    \fig{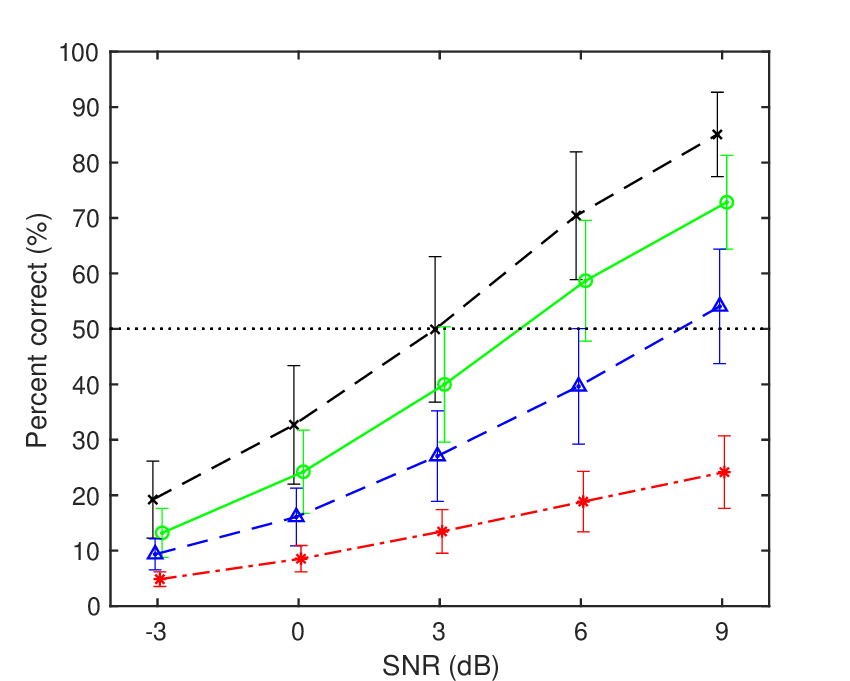}{0.33\linewidth}
    {\vspace{-10pt} \label{fig:GESI_mis_Lab_OptmMean_rPwr60} (e) GESI ($\rho = 0.60$)}
    \fig{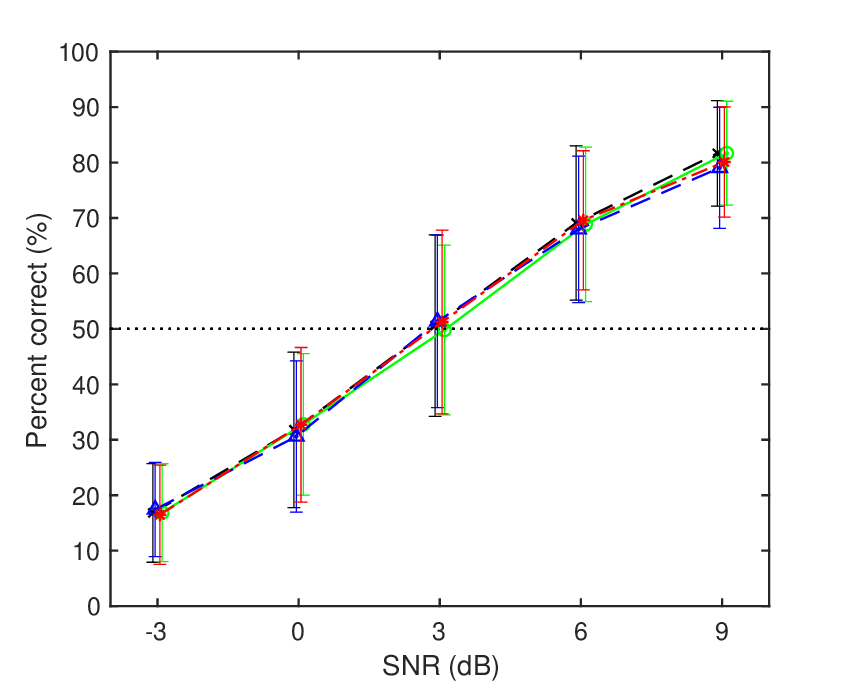}{0.33\linewidth}{\vspace{-10pt}\label{fig:ESTOI_mis_Lab_OptmMean} (f) ESTOI }
    }

    \vspace{-12pt}

    \figline{
    \fig{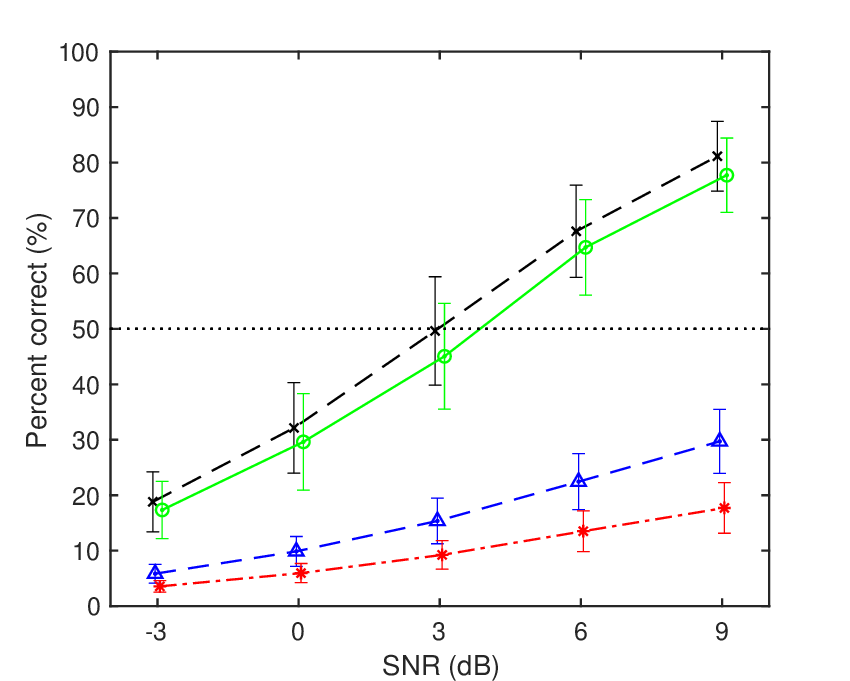}{0.33\linewidth}{\vspace{-10pt}\label{fig:HASPI_mis_Lab_OptmMean} (g) HASPIv1}
    \fig{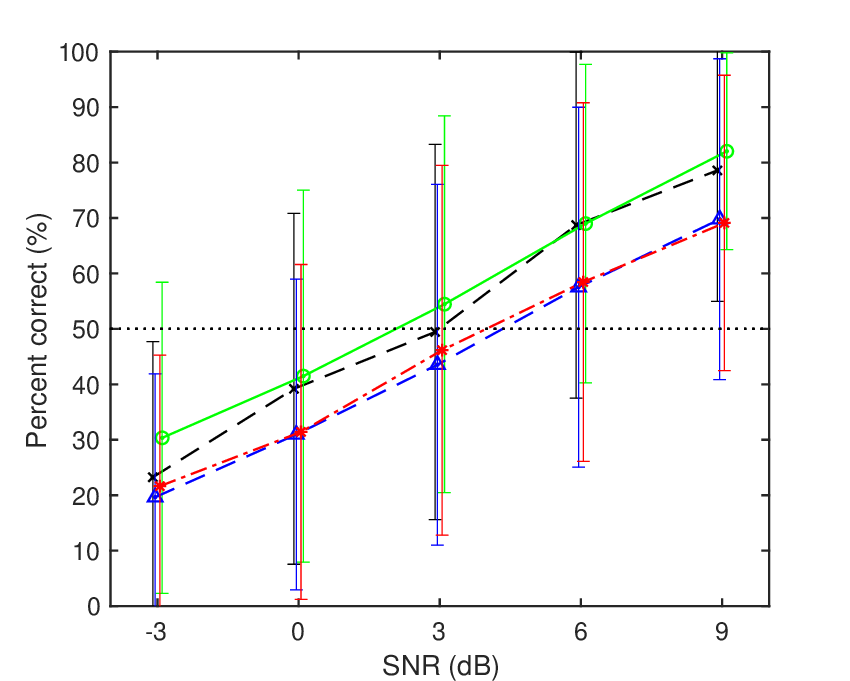}{0.33\linewidth}{ \vspace{-10pt}\label{fig:HASPIv2_mis_Lab_OptmMean} (h) HASPIv2}
    \fig{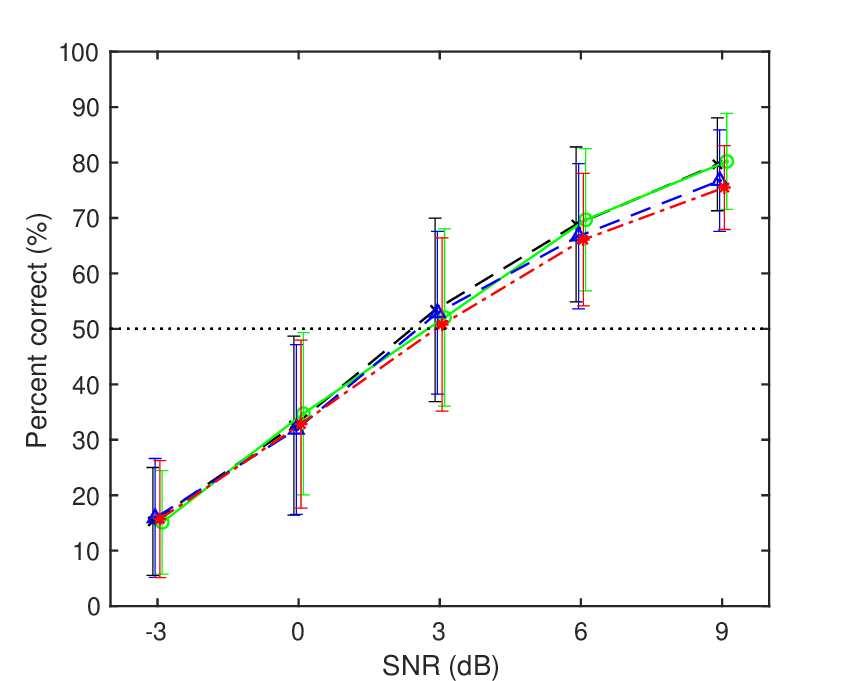}{0.33\linewidth}{ \vspace{-10pt}\label{fig:MBSTOI_mis_Lab_OptmMean} (i) MBSTOI}
    }


    \caption{ 
    \label{fig:PredOIM_Eval1}
    SI prediction results in Eval.1. For comparison, human subjective results on male speech experiments for laboratory (a) and remote (b), which are exactly the same as Figs.~\ref{fig:RsltSbj}(a) and \ref{fig:RsltSbj}(b), are reproduced here. SI predictions by STOI (c), GESI ($\rho=0.55$) (d), GESI ($\rho=0.60$) (e), ESTOI (f), HASPIv1 (g), HASPIv2 (h), MBSTOI (i). The mean value and standard deviation (SD) across the participants and words. }

\end{figure*}


\paragraph{Results} 
Figures~\ref{fig:PredOIM_Eval1}(d) and \ref{fig:PredOIM_Eval1}(e) 
show the SI values predicted by GESI for the
laboratory and remote experiments when the $\rho$ values were set to 0.55 and 0.60
as listed in Table \ref{tab:Npip_rho}. 
The results are very similar to the human subjective SI values shown in panels (a) and (b).
This means that the SI values of the low-level condition can be reasonably predicted by adjusting the $\rho$ value. It was also possible to reasonably predict the 70-yr and 80-yr conditions. 

Figure~\ref{fig:PredOIM_Eval1}(g) 
shows the prediction results using HASPIv1.
The SI values of the low-level condition were just below the unprocessed line. There is no way to introduce the listening condition dependence described in sections \ref{sec:SRTvsTonePiP_WHIS} and \ref{sec:Predict_Challenge}. Therefore, it was not possible to consistently predict the difference in subjective SI values for the low-level conditions in the laboratory and remote experiments shown in
panels (a) and (b). 
Moreover, the prediction results on 70-yr and 80-yr were much smaller than those observed in the experimental results.

Figure~\ref{fig:PredOIM_Eval1}(h)
shows the prediction results using HASPIv2. The standard deviations of the SI values were much larger for HASPIv2 than for the other OIMs.
There is almost no difference between the 70-yr and 80-yr conditions. The results were much worse than those in HASPIv1.
This is probably because the NN of HASPIv2 was trained with English sentence databases as described in section~\ref{sec:Eval_Obj_HASPIv2} and was not intended to predict the word SI. It should be noted that SI values vary widely depending on speech material and cognitive factors. Therefore, the current version of the NN output of HASPIv2 may be better at predicting sentence SI, but does not seem suitable for evaluating SE algorithms that aim to improve the SI of phonemes, syllables, or words.
We do not use it in the following evaluation.

Figures~\ref{fig:PredOIM_Eval1}(c), \ref{fig:PredOIM_Eval1}(f), and \ref{fig:PredOIM_Eval1}(i)
show the prediction results using STOI, ESTOI, and MBSTOI with diotic input. 
The SI values could not be predicted as expected. The input levels of the reference and test sounds in STOI and ESTOI were normalized to the same level. Although MBSTOI allows for different input levels, the Pearson correlation used in it normalizes the distribution so that the effect is similar to that of the input level normalization. Therefore, they are not suitable for the development of hearing assistive devices that require the evaluation of different input levels.



\subsection{\textbf{Eval.2}: Prediction of the average SI without using simulated HL sounds with limited parameter value settings}
\label{sec:Eval_WithoutWHIS}

As described in section \ref{sec:Predict_motivation}, the main goal of GESI is to develop an SI measure for HI listeners. Therefore, it is important to predict the SI values of 70-yr and 80-yr conditions without using simulated HL sounds. 
Furthermore, it is desirable to predict all subjective SI values in Fig.~\ref{fig:RsltSbj}, regardless of the type of experiment (male, female, laboratory, or remote), with a single set of sigmoid parameters in Eqs.~\ref{eq:STOIGESI_sigmoid} and \ref{eq:HASPI_sigmoid}, determined from a small subset of the results.
In this evaluation, we set the sigmoid parameters exactly the same as those used in Eval.1, which were determined only from the unprocessed condition of the male speech laboratory experiment in Fig.~\ref{fig:RsltSbj}(a). 
We used HASPIv1 for comparison simply because it was better than HASPIv2 in Eval.1, as described in section~\ref{sec:Eval_WithWHIS}.

\subsubsection{Prediction by GESI}
\label{sec:Eval_WithoutWHIS_GESI}

\paragraph{Settings} GCFB in the first stage of GESI can simulate cochlear outputs or excitation patterns~\citep{moore2013introduction} of an HI listener~\citep{irino2020gammachirp,irino2023hearing}. We set the average hearing levels of 70-yr and 80-yr shown in 
Table~\ref{tab:HL} and the compression health to $\alpha=0.5$, which corresponds to a moderate dysfunction of the active cochlear mechanism. It is important to set the compression health $\alpha$ independently of the audiogram because the ratio of active to passive hearing loss can vary from one HI listener to another.
For predicting the SI of the low-level condition, we set the parameter of GCFB as an NH setting (i.e., the hearing level of 0\,dB and $\alpha=1$), and the level of the test signal was simply reduced to -20\,dB. This setting reflects the situation of the subjective experiments.
The parameters $a$ and $b$ were the same as those in Eval.1, as shown in Table ~\ref{tab:Eval_ParamSet_Eval12}.
The $\rho$ values were set individually for each listener.



\begin{figure*}[t] 

    \parskip=0pt
    \baselineskip=0pt

\figline{
\fig{Fig3_ExpRsltPC_WHIS_Lab.eps}{0.24\linewidth}{\vspace{-10pt}\label{fig:RsltSbj_mis_Lab_Eval2} (a) Human: Male, Lab.}
\fig{Fig3_ExpRsltPC_WHIS_Remote.eps}{0.24\linewidth}{ \vspace{-10pt}\label{fig:RsltSbj_mis_Remote_Eval2} (b) Human: Male, Remote  }

\fig{Fig3_ExpRsltPC_WHISf_Lab.eps}{0.24\linewidth}{\vspace{-10pt}\label{fig:RsltSbj_fhi_Lab_Eval2} (c) Human: Female, Lab.}
\fig{Fig3_ExpRsltPC_WHISf_Remote.eps}{0.24\linewidth}{\vspace{-10pt}\label{fig:RsltSbj_fhi_Remote_Eval2} (d) Human: Female, Remote}
}
    \vspace{-12pt}
    
    \figline{
    \fig{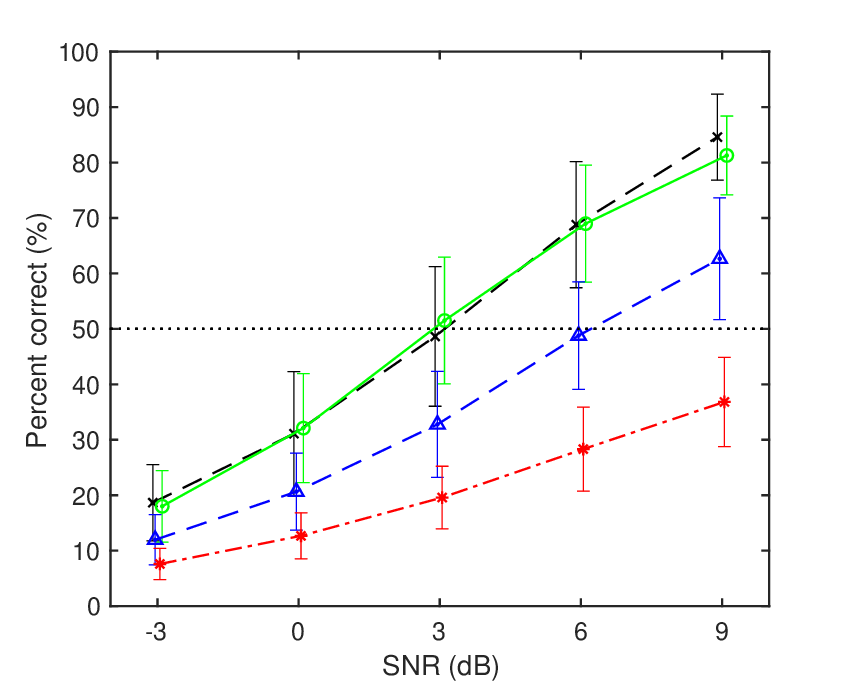}{0.24\linewidth}
    {\vspace{-10pt}\label{fig:GESI_mis_Lab_FixedSigmoid} (e) GESI: Male, Lab.}
    \fig{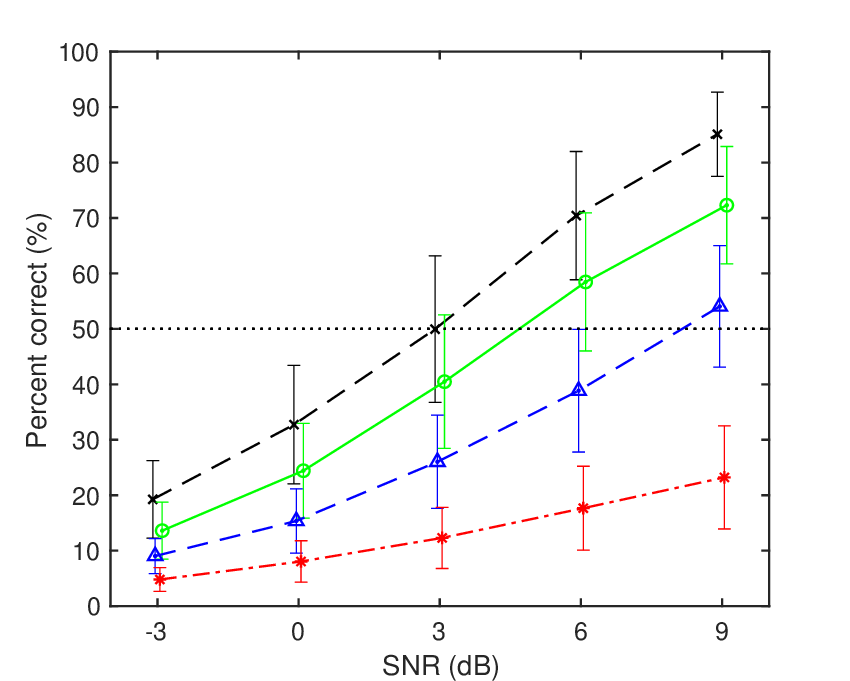}{0.24\linewidth}
{\vspace{-10pt}\label{fig:GESI_mis_Remote_FixedSigmoid} (f) GESI: Male, Remote}


    \fig{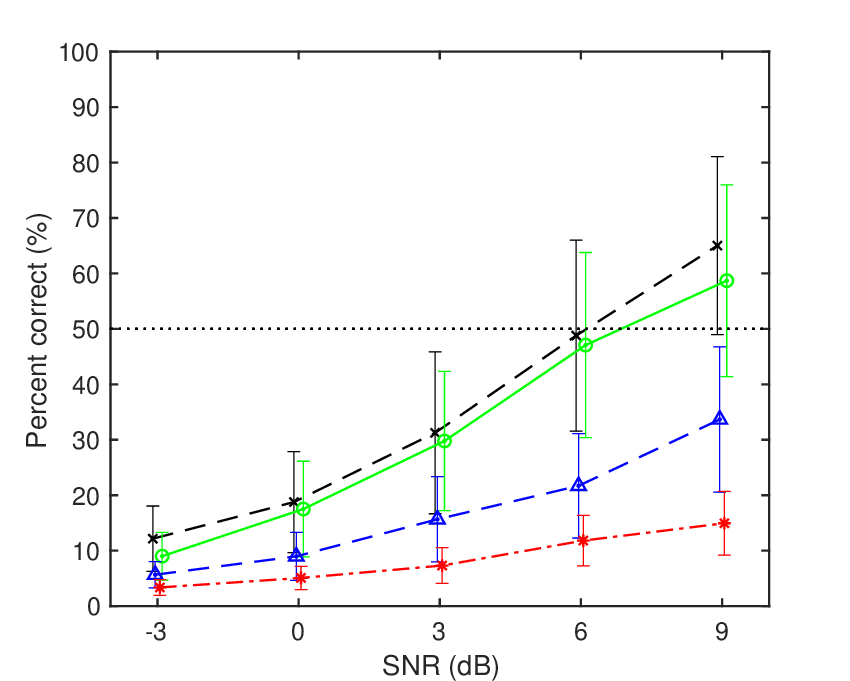}{0.24\linewidth}{\vspace{-10pt}\label{fig:GESI_fhi_Lab_FixedSigmoid} (g) GESI: Female, Lab.}
    \fig{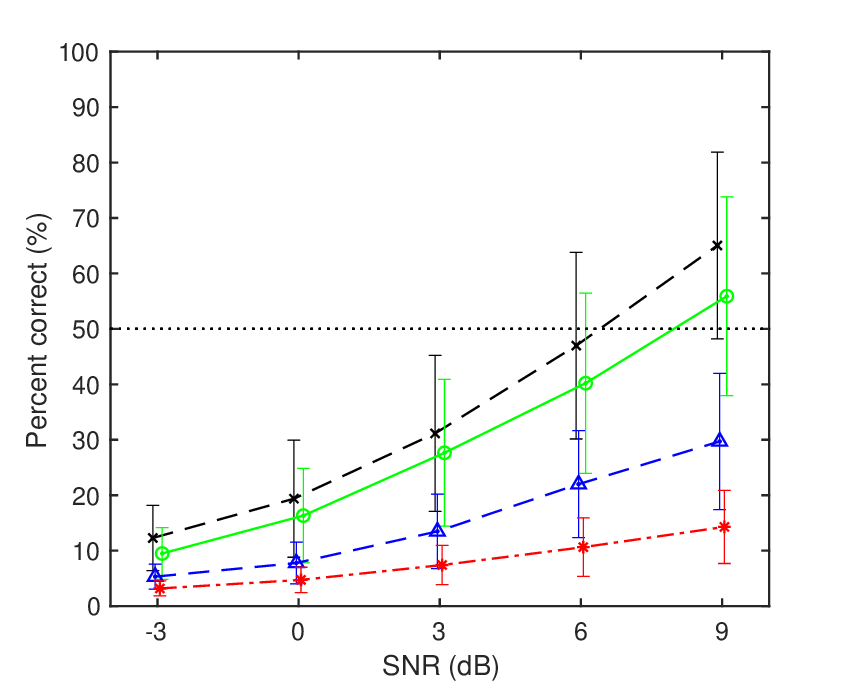}{0.24\linewidth}{ \vspace{-10pt}\label{fig:GESI_fhi_Remote_FixedSigmoid} (h) GESI: Female, Remote}
    }

    \vspace{-12pt}

    \figline{
    \fig{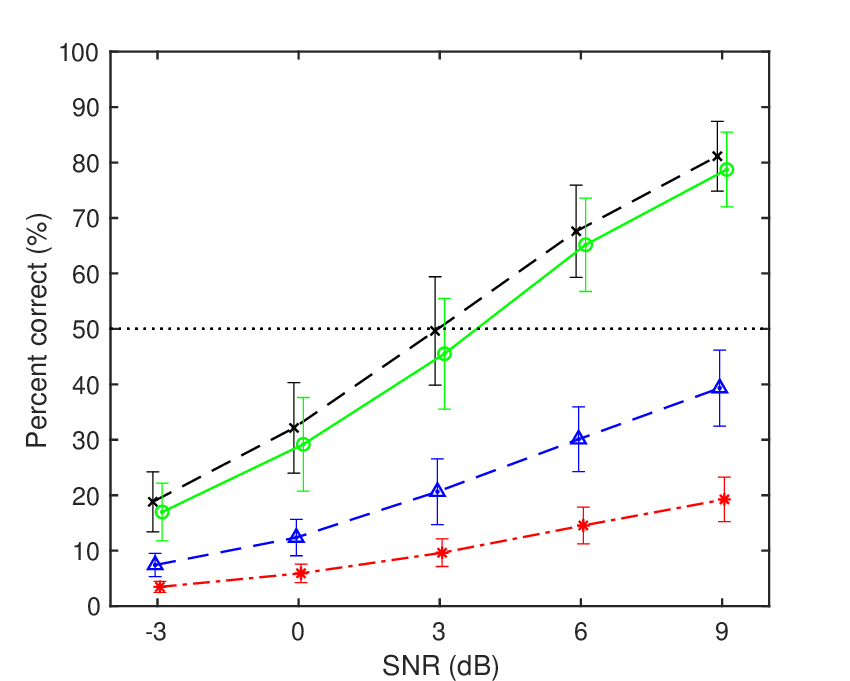}{0.24\linewidth}{\vspace{-10pt}\label{fig:HASPI_mis_Lab_FixedSigmoid} (i) HASPIv1: Male, Lab.}
    \fig{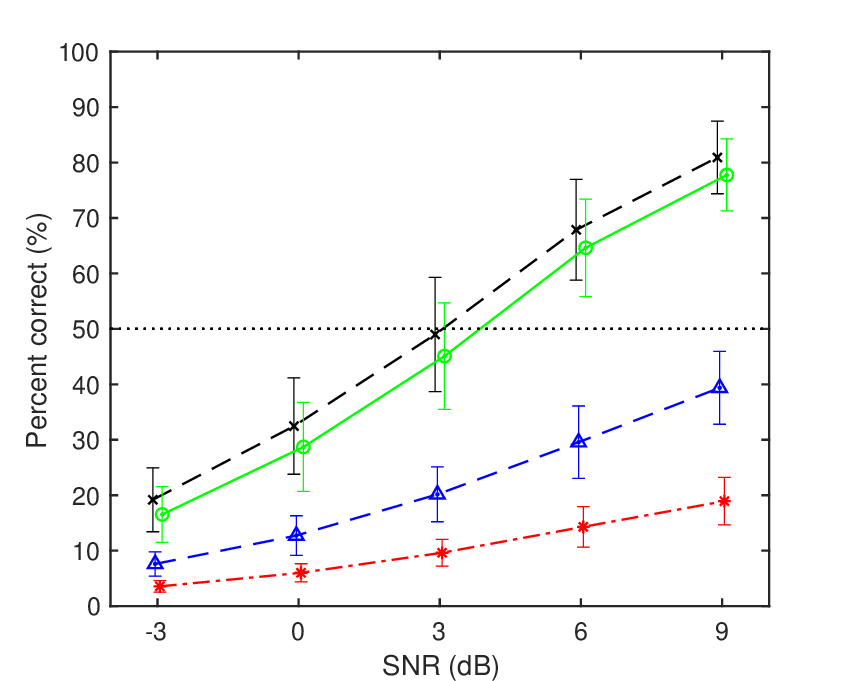}{0.24\linewidth}{ \vspace{-10pt}\label{fig:HASPI_mis_Remote_FixedSigmoid} (j) HASPIv1: Male, Remote}
   


    \fig{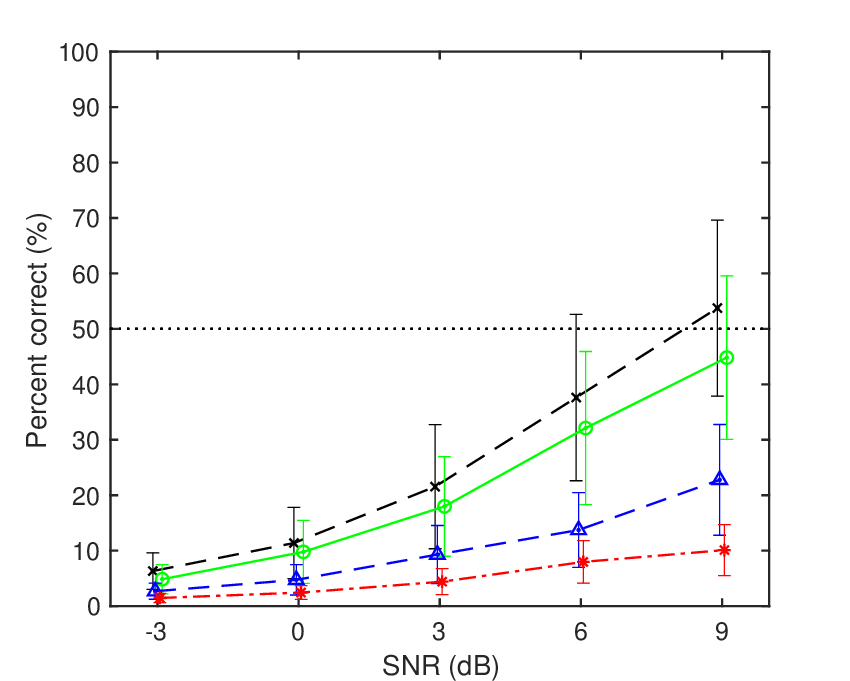}{0.24\linewidth}{\vspace{-10pt}\label{fig:HASPI_fhi_Lab_FixedSigmoid} (k) HASPIv1: Female, Lab.}
    \fig{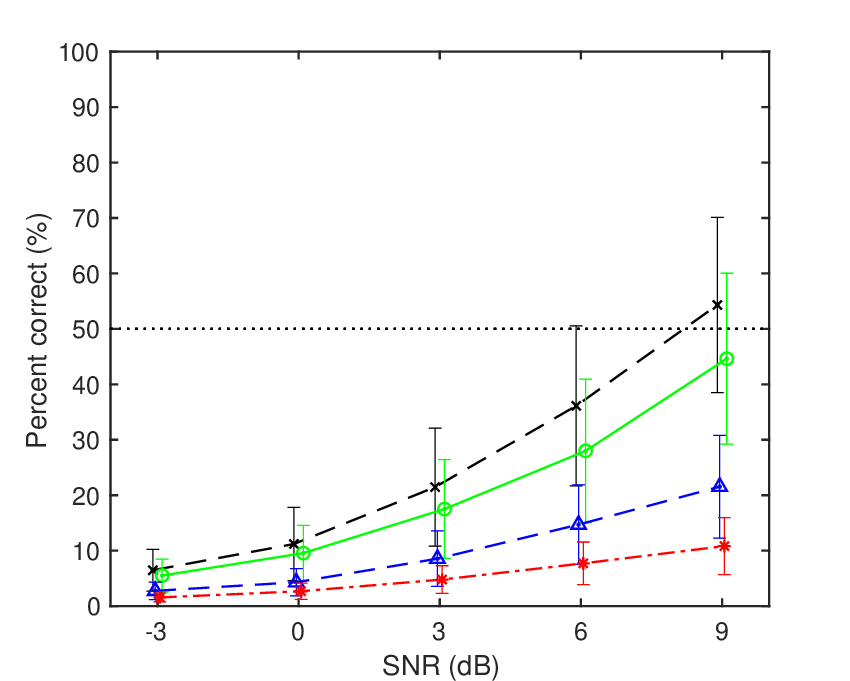}{0.24\linewidth}{ \vspace{-10pt}\label{fig:HASPI_fhi_Remote_FixedSigmoid} (l) HASPIv1: Female, Remote}
    }

    \caption{\label{fig:OIM_FixedSigmoid} 
        SI prediction results in Eval.2 using GESI and HASPIv1 without using WHIS sounds. For comparison, human subjective results on male laboratory (a), male remote (b), female laboratory (c), and female remote (d) experiments are reproduced from Figs.~\ref{fig:RsltSbj}(a), \ref{fig:RsltSbj}(b), \ref{fig:RsltSbj}(b), and \ref{fig:RsltSbj}(d). 
        SI predictions by GESI are
        shown in the middle row ((e), (f), (g), and (h)) and HASPIv1 are shown in the bottom row ((i), (j), (k), and (l)). The mean value and standard deviation (SD) across the participants and words. }
    

\end{figure*}

\paragraph{Results} 
The middle four panels (e), (f), (g), and (h) in Fig.~\ref{fig:OIM_FixedSigmoid} show the SI values predicted by GESI corresponding to the four subjective SI experiments shown in the top panels (a), (b), (c), and (d).
The differences between the unprocessed and low-level conditions in the male speech experiments shown in panels (a) and (b) were correctly reproduced. The SI values were predicted to be lower in the female speech experiments than in the male speech experiments.
The SI values of 70-yr and 80-yr conditions for both the male and female speech experiments were fairly well predicted without using WHIS sounds.
The quantitative evaluation is described later in section \ref{sec:Eval_WithoutWHIS_ComparePredErr}.

\subsubsection{Prediction by HASPIv1}
\label{sec:Eval_WithoutWHIS_HASPI}

\paragraph{Settings} 
HASPIv1~\citep{kates2014hearing} requires hearing levels between 250\,Hz and 6000\,Hz. So, we set the 70-yr and 80-yr hearing levels, as shown in Table~\ref{tab:HL} with an interpolated value at 6000\,Hz because there was no hearing level data at that frequency~\citep{tsuiki2002nihon}. 
There is no additional input parameter for the HL simulation, such as the compression health $\alpha$ in GESI. The degree of the active HL corresponding to $\alpha$ is determined automatically from the audiogram. 
The parameters $B$, $C$, and $A_{high}$ were the same as those in Eval.1, as shown in Table ~\ref{tab:Eval_ParamSet_Eval12}.

\paragraph{Results} The bottom four panels (i), (j), (k), and (l) in Fig.~\ref{fig:OIM_FixedSigmoid} 
are the SI values predicted by HASPIv1 corresponding to the four subjective SI experiments shown in the top panels (a), (b), (c), and (d).
The most notable point is that the SI values in the laboratory and remote experiments were virtually the same. Therefore, it was not possible to reproduce the differences between the unprocessed and low-level conditions shown in panels (a) and (b). This is mainly because HASPIv1 does not provide parameters to reflect the different listening conditions, such as $\rho$ in GESI.
The predicted SI values in the female speech experiments shown in panels (k) and (l) were predicted to be much lower than the subjective SI values in panels (c) and (d). 
This implies that the generalization ability of HASPIv1 is lower than that of GESI.

\begin{table}[t]
\centering
    \caption{
    Mean RMS errors in the SI prediction Eval.2. The RMS error between the subjective and predicted SI values across 5 SNRs was calculated for each subject and each HL condition.
    The mean RMS error across the subjects is shown in the middle rows. The {\it t}-test was performed on the mean RMS errors between GESI and HASPIv1 (*$p<0.05$, **$p<0.01$, ***$p<0.001$). The bottom row shows the correlation coefficients between all subjective and predicted SI values. The $p$ values were much smaller than 0.001 for all coefficients.
    Bold type indicates better results, i.e.~lower RMS error or higher correlation coefficient.     
}   \label{tab:RMSECorr_WHIS_FixedSigmoid}

     \begin{tabular}{lrrrrr}
        \toprule
        Male speech &\multicolumn{2}{c}{Laboratory} & &\multicolumn{2}{c}{Remote}\\
        \cline{2-3}\cline{5-6}
        &\multicolumn{1}{c}{GESI} &\multicolumn{1}{c}{HASPIv1} & &\multicolumn{1}{c}{GESI} &\multicolumn{1}{c}{HASPIv1}\\
        \midrule
        Mean RMS error & & & & &\\
        ~ Unprocessed &\textbf{9.98} &10.22 & &\textbf{13.38} &13.57 \\
        ~ Low-level &\textbf{10.45} &10.97 & &\textbf{12.96}$^{\footnotesize{*}}$ &15.57 \\
        ~ 70-yr, $\alpha=0.5$ &\textbf{10.02} &12.10 & &\textbf{10.22}$^{\footnotesize{*}}$ &12.75 \\
        ~ 80-yr, $\alpha=0.5$ &\textbf{10.27}$^{\footnotesize{***}}$ &16.59 & &\textbf{10.66}$^{\footnotesize{**}}$ &13.13 \\
        \midrule
        Corr. Coef. &\textbf{0.91} &0.88 & &\textbf{0.88} &0.82\\
        \bottomrule
    \end{tabular}

\vspace{10pt}
\centering
    \begin{tabular}{lrrrrr}
        \toprule
        Female speech &\multicolumn{2}{c}{Laboratory} & &\multicolumn{2}{c}{Remote}\\
        \cline{2-3}\cline{5-6}
        &\multicolumn{1}{c}{GESI} &\multicolumn{1}{c}{HASPIv1} & &\multicolumn{1}{c}{GESI} &\multicolumn{1}{c}{HASPIv1}\\
        \midrule
        Mean RMS error & & & & &\\
        ~ Unprocessed  &\textbf{13.32}$^{\footnotesize{***}}$ &20.09 & &\textbf{13.90}$^{\footnotesize{***}}$ &20.44 \\
        ~ Low-level  &\textbf{15.40}$^{\footnotesize{***}}$ &24.85 & &\textbf{15.87}$^{\footnotesize{***}}$ &23.60 \\
        ~ 70-yr, $\alpha=0.5$ &\textbf{7.42} & 9.15 & &\textbf{9.51}$^{\footnotesize{***}}$ &12.08 \\
        ~ 80-yr, $\alpha=0.5$ &\textbf{7.40}$^{\footnotesize{***}}$ &9.56 & &\textbf{9.92}$^{\footnotesize{***}}$ &11.61 \\
        \midrule
        Corr. Coef. &\textbf{0.90} &0.88 & &\textbf{0.89} &0.86 \\
        \bottomrule
    \end{tabular}
\end{table}

\subsubsection{Prediction error and correlation}
\label{sec:Eval_WithoutWHIS_ComparePredErr}

We quantitatively evaluated the predictability of GESI and HASPIv1. 
Table \ref{tab:RMSECorr_WHIS_FixedSigmoid} shows the RMS error and correlation coefficients between the subjective and predicted SI values for the male (top) and female (bottom) speech experiments. Bold type indicates better results, i.e.~lower RMS error or higher correlation coefficient.
The top four rows of each table show the RMS errors obtained for each HL condition, and the bottom two rows show the correlation coefficient for all SI values and its $p$ value.
The RMS errors of GESI were always smaller than those of HASPIv1, without any exception.
The {\it t}-test was performed on the mean RMS errors for GESI and HASPIv1 in each HL condition. Significant differences were found for eleven of the sixteen combinations. In particular, the RMS errors were almost always significantly smaller in the female speech experiments.
The generalization ability across the different speech (male versus female) is higher in GESI than in HASPIv1.
The bottom two rows show the correlation coefficients between the subjective and predicted SI values for all data points.
The correlation coefficients were always higher for GESI than for HASPIv1. 





\begin{figure*}[t]

    \parskip=0pt
    \baselineskip=0pt

\figline{
\fig{Fig3_ExpRsltPC_WHIS_Lab.eps}{0.24\linewidth}{\vspace{-10pt}\label{fig:RsltSbj_mis_Lab_Eval3} (a) Human: Male, Lab.}
\fig{Fig3_ExpRsltPC_WHIS_Remote.eps}{0.24\linewidth}{ \vspace{-10pt}\label{fig:RsltSbj_mis_Remote_Eval3} (b) Human: Male, Remote  }

\fig{Fig3_ExpRsltPC_WHISf_Lab.eps}{0.24\linewidth}{\vspace{-10pt}\label{fig:RsltSbj_fhi_Lab_Eval3} (c) Human: Female, Lab.}
\fig{Fig3_ExpRsltPC_WHISf_Remote.eps}{0.24\linewidth}{\vspace{-10pt}\label{fig:RsltSbj_fhi_Remote_Eval3} (d) Human: Female, Remote}
}
    \vspace{-12pt}
    
    \figline{
    \fig{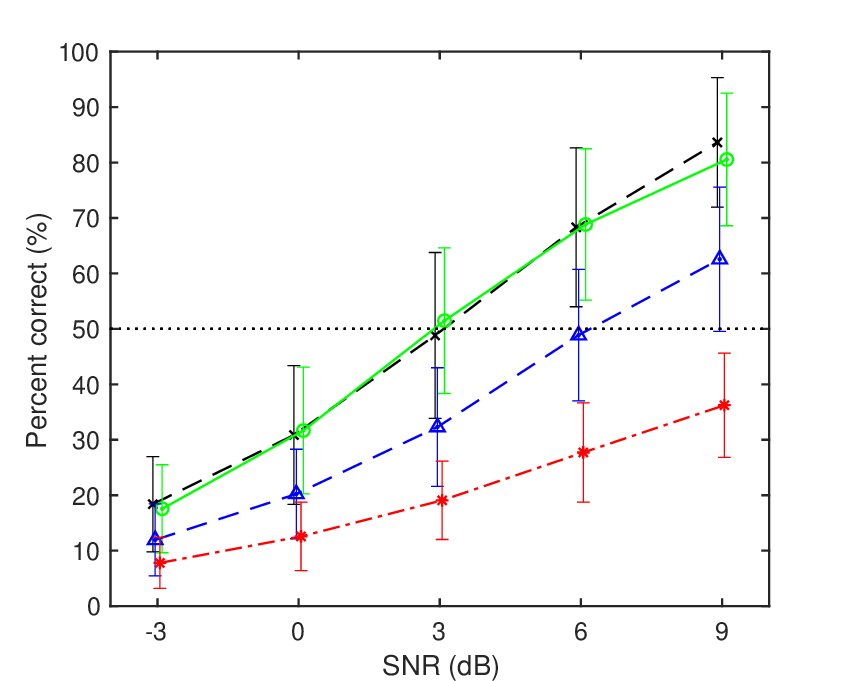}{0.24\linewidth}
    {\vspace{-10pt}\label{fig:GESI_mis_Lab_OptmSbj} (e) GESI: Male, Lab.}
    \fig{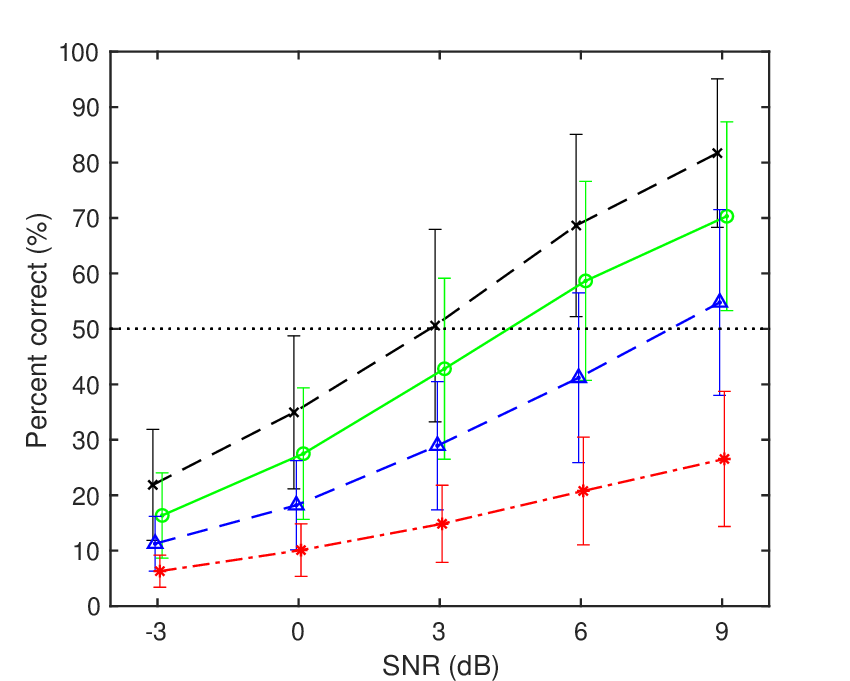}{0.24\linewidth}
    { \vspace{-10pt}\label{fig:GESI_mis_Remote_OptmSbj} (f) GESI: Male, Remote}


    \fig{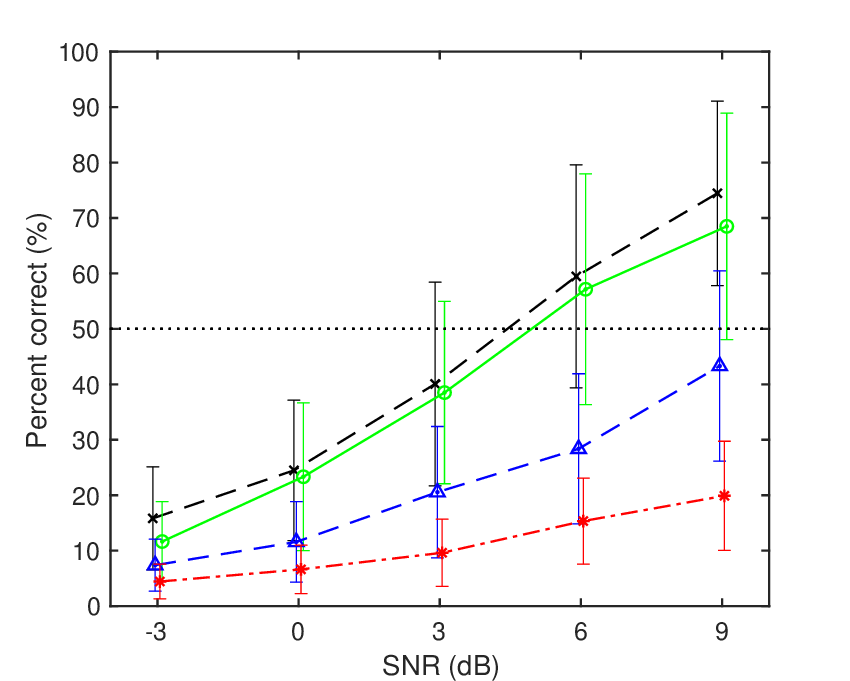}{0.24\linewidth}{\vspace{-10pt}\label{fig:GESI_fhi_Lab_OptmSbj} (g) GESI: Female, Lab.}
    \fig{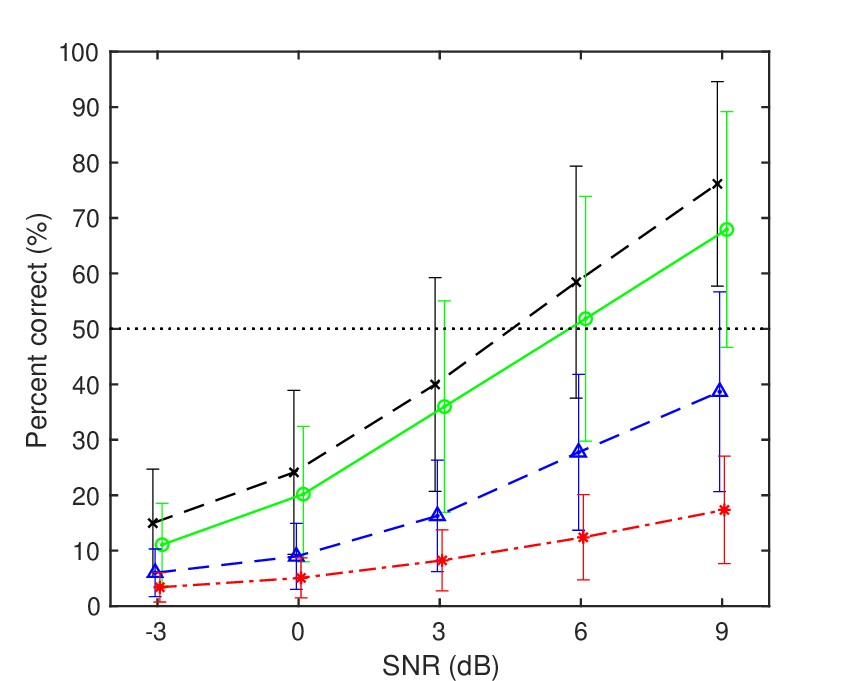}{0.24\linewidth}{ \vspace{-10pt}\label{fig:GESI_fhi_Remote_OptmSbj} (h) GESI: Female, Remote}
    }

    \vspace{-12pt}

    \figline{
    \fig{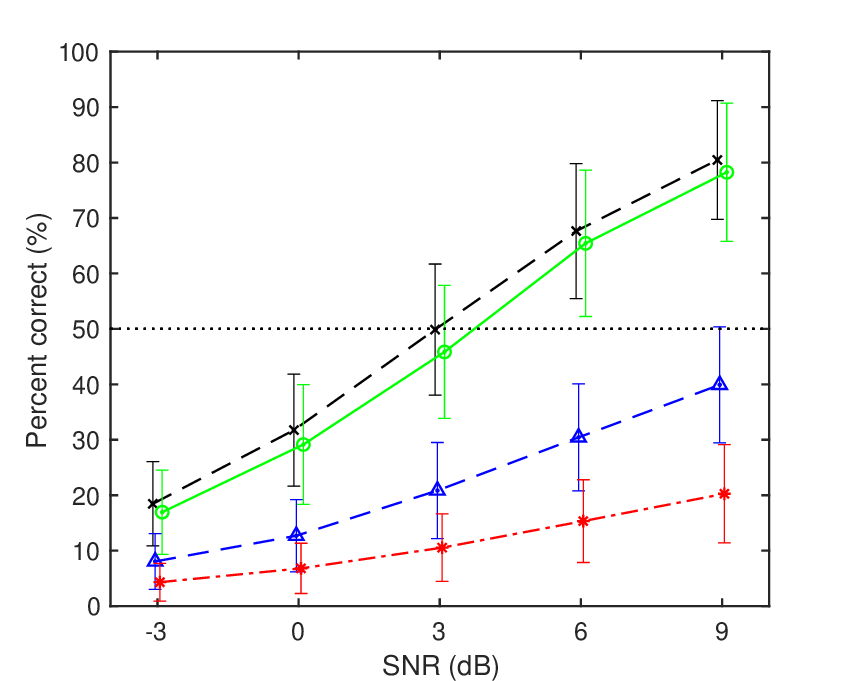}{0.24\linewidth}{\vspace{-10pt}\label{fig:HASPI_mis_Lab_OptmSbj} (i) HASPIv1: Male, Lab.}
    \fig{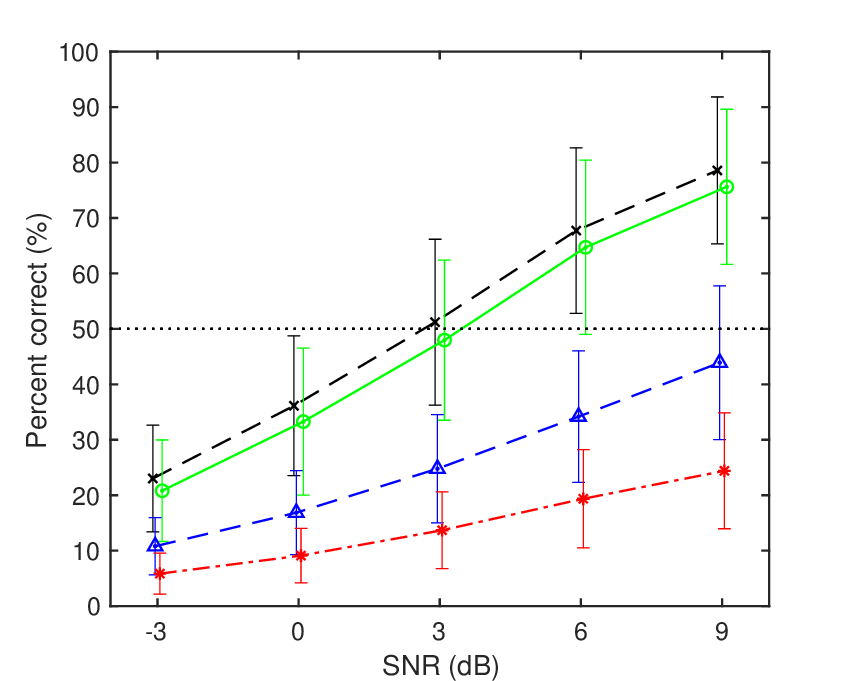}{0.24\linewidth}{ \vspace{-10pt}\label{fig:HASPI_mis_Remote_OptmSbj} (j) HASPIv1: Male, Remote}


    \fig{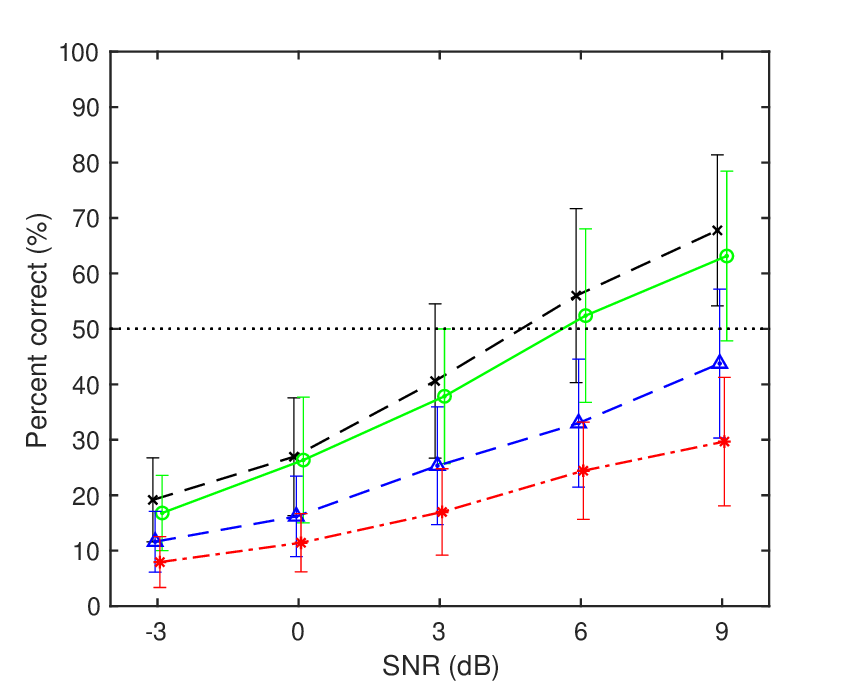}{0.24\linewidth}{\vspace{-10pt}\label{fig:HASPI_fhi_Lab_OptmSbj} (k) HASPIv1: Female, Lab.}
    \fig{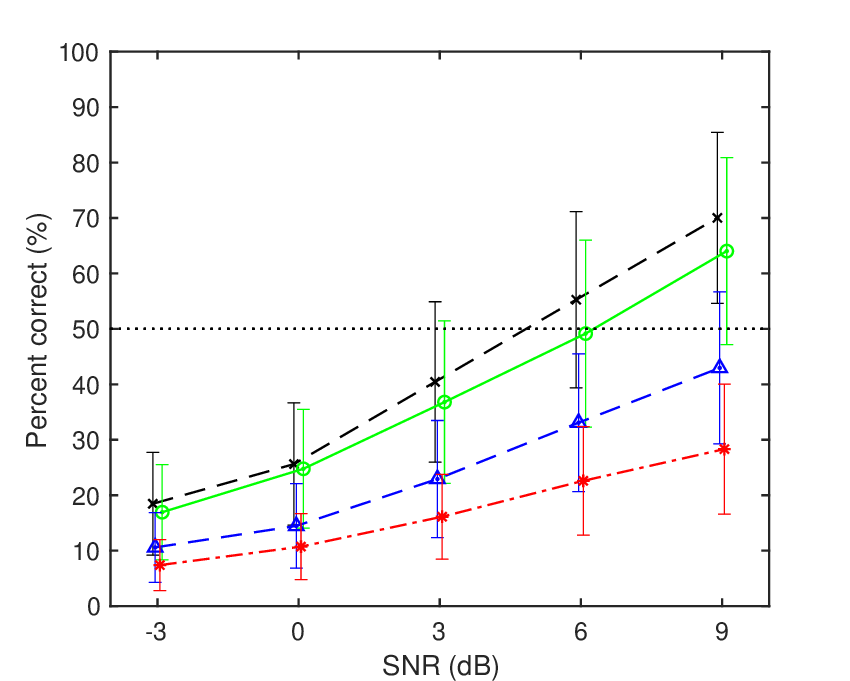}{0.24\linewidth}{ \vspace{-10pt}\label{fig:HASPI_fhi_Remote_OptmSbj} (l) HASPIv1: Female, Remote}
    }


    \caption{\label{fig:OIM_OptmSbj}
        SI prediction results in Eval.3 using GESI and HASPIv1 without using WHIS sounds. For comparison, human subjective results on male laboratory (a), male remote (b), female laboratory (c), and female remote (d) experiments are reproduced from Figs.~\ref{fig:RsltSbj}(a), \ref{fig:RsltSbj}(b), \ref{fig:RsltSbj}(b), and \ref{fig:RsltSbj}(d). 
        SI predictions by GESI are
        shown in the middle row ((e), (f), (g), and (h)) and HASPIv1 are shown in the bottom row ((i), (j), (k), and (l)).  The mean value and standard deviation (SD) across the participants and words. 
    }

\end{figure*}



\subsection{\textbf{Eval.3}: Prediction of the SI of the individual listeners}
\label{sec:Eval_Indiv}
The purpose of this section was to evaluate GESI and HASPIv1 in predicting the SI values for individual listeners. It is important to predict the individual SI values in different situations from a small number of SI tests without using WHIS sounds.


\subsubsection{Prediction by GESI}
\label{sec:Eval_Indiv_GESI}
\paragraph{Settings} 
The parameters were set individually for each participant.
The setting of HL of 70-yr and 80-yr in GCFB is the same as described in section \ref{sec:Eval_WithoutWHIS_GESI}.
The $\rho$ value for the four experiments were calculated using Eq.~\ref{eq:rho=05} from the average reported number, $\bar N_{pip}$, of each listener.
The coefficients $a$ and $b$ in the SI sigmoid function in Eq.~\ref{eq:STOIGESI_sigmoid} were determined by least squares for each participant only from their subjective SI values (five SNR points) of the unprocessed condition. 
Therefore, the values of $a$ and $b$ were individually different. This is in contrast to the parameter settings in the previous sections.
Then, the SI values of all HL conditions were predicted for the male and female speech experiments. The results are presented with means and standard deviations across participants.

\paragraph{Results} The middle four panels (e), (f), (g), and (h) in Fig.~\ref{fig:OIM_OptmSbj} 
show the SI values predicted by GESI corresponding to the four subjective SI experiments shown in the top panels (a), (b), (c), and (d).
Although the SI values in the 70-yr condition were overestimated by about 10\% points at SNR=9\,dB, the other SI values were quite well predicted. The differences between the unprocessed and low-level conditions in the male speech experiments shown in panels (a) and (b) were again correctly reproduced in this case.



\subsubsection{Prediction by HASPIv1}
\label{sec:Eval_Indiv_HASPI}

\paragraph{Settings} 
The HL parameters were set similarly as described in section \ref{sec:Eval_WithoutWHIS_HASPI}. 
The coefficients, $B$, $C$, and $A_{high}$, were also determined by least squares for each participant only from the subjective SI values (five SNR points) of the unprocessed condition of the male speech laboratory experiment. Therefore, the values were individually different. The SI values of all HL conditions were then predicted for the male and female speech experiments.

\paragraph{Results} 
The bottom four panels (i), (j), (k), and (l) in Fig.~\ref{fig:OIM_OptmSbj} 
show the the SI values predicted by HASPIv1 corresponding to the four subjective SI experiments shown in the top panels (a), (b), (c), and (d).
The SI values in the laboratory and remote experiments were again virtually the same, as observed in Fig.~\ref{fig:OIM_FixedSigmoid}. 
Furthermore, the SI values in the 70-yr and 80-yr conditions were almost the same regardless of the type of experiment (laboratory vs.\ remote and male vs.\ female). These results suggest that HASPIv1 did not predict the SI values on an individual basis.
One possible reason is that the coefficients of $B$, $C$, and $A_{high}$, which correspond to the cepstral correlations ($c$) and the auditory coherence ($a_{high}$),
were not sufficiently estimated from the five subjective SI values alone, although the number of data points is not theoretically insufficient. 
Another possible reason is that the use of $c$ and $a_{high}$ is not appropriate for this speech-in-noise condition. 
In this case, and when using $a_{low}$ and $a_{mid}$ in addition, more data points may be needed for a reasonable estimate because these features have been linearly combined within a sigmoid function as formulated in Eqs.~\ref{eq:HASPI_sigmoid} and \ref{eq:HASPI_linearAdd}.
This is in contrast to GESI, which uses a single metric $d$ for the SI sigmoid function in Eq.~\ref{eq:STOIGESI_sigmoid}.
HASPIv1 was designed to predict the SI averaged across listeners with similar audiograms, not the performance of individual subjects ~\citep{kates2014hearing}, which may be another reason for the difference in performance.

\begin{table}[t] 
\caption{Mean RMS errors in the SI prediction Eval.3. The RMS error between the subjective and predicted SI values across 5 SNRs was calculated for each subject and each HL condition.
The mean RMS error across the subjects is shown in the middle rows. The {\it t}-test was performed on the mean RMS errors between GESI and HASPIv1 (*$p<0.05$, **$p<0.01$, ***$p<0.001$). The bottom row shows the correlation coefficients between all subjective and predicted SI values. The $p$ values were much smaller than 0.001 for all coefficients.
Bold type indicates better results, i.e.~lower RMS error or higher correlation coefficient.
} 
\label{tab:RMSECorr_WHIS_OptmSbj}
\centering
\begin{tabular}{lrrrrr}
    \toprule
    Male speech & \multicolumn{2}{c}{ Laboratory}	&&	\multicolumn{2}{c}{Remote}\\
    \cline{2-3}\cline{5-6}
    &\multicolumn{1}{c}{GESI} & \multicolumn{1}{c}{HASPIv1} &&  \multicolumn{1}{c}{GESI} & \multicolumn{1}{c}{HASPIv1}\\
    \midrule
    Mean RMS error &&&&&\\
    ~ Unprocessed & 7.06 &\textbf{7.02} & &\textbf{8.03} &8.16\\
    ~ Low-level  & \textbf{11.94} &12.52 & &\textbf{12.83}$^{\footnotesize{**}}$ &15.97\\
    ~ 70-yr, $\alpha=0.5$ & \textbf{10.80} &12.89 & &\textbf{12.17} &14.12\\
    ~ 80-yr, $\alpha=0.5$ & \textbf{10.55}$^{\footnotesize{**}}$ &16.24 & &\textbf{12.28} &14.27\\
    \midrule
    Corr Coef.& \textbf{0.91} &0.88 & &\textbf{0.88} &0.83\\
    \bottomrule
\end{tabular}

\vspace{10pt}
\centering
\begin{tabular}{lrrrrr}
    \toprule
    Female speech& \multicolumn{2}{c}{Laboratory} &&	\multicolumn{2}{c}{Remote}\\
    \cline{2-3}\cline{5-6}
    & \multicolumn{1}{c}{GESI} & \multicolumn{1}{c}{HASPIv1} && \multicolumn{1}{c}{GESI} & \multicolumn{1}{c}{HASPIv1}\\
    \midrule
    Mean RMS error &&&&&\\
    ~ Unprocessed	       & \textbf{8.75}$^{\footnotesize{**}}$ &10.03 & &\textbf{8.18}$^{\footnotesize{***}}$ &9.62 \\
    ~ Low-level  & \textbf{12.78} &13.47 & &\textbf{13.08} &13.57 \\
    ~ 70-yr, $\alpha=0.5$ & \textbf{10.77}$^{\footnotesize{**}}$ &13.11 & &\textbf{10.50}$^{\footnotesize{***}}$ &13.97 \\
    ~ 80-yr, $\alpha=0.5$ & \textbf{8.31} &10.99 & &\textbf{9.90} &11.04 \\
    \midrule
    Corr Coef. & \textbf{0.89} &0.88 & &\textbf{0.89} &0.86 \\
    \bottomrule
\end{tabular}
\end{table}

\subsubsection{Prediction error and correlation}

We statistically evaluated the predictability of GESI and HASPIv1. 
Table \ref{tab:RMSECorr_WHIS_OptmSbj} shows the RMS error and correlation coefficients between the subjective and predicted SI values for the male (top) and female (bottom) speech experiments. 
The top four rows of each table show the RMS errors obtained for each HL condition, and the bottom two rows show the correlation coefficient for all SI values and its $p$ value. Bold type indicates better results, i.e.~lower RMS error or higher correlation coefficient.
The RMS errors of GESI were always smaller than those of HASPIv1, except for the unprocessed condition in the male speech laboratory experiment, which is closed data for parameter setting. The {\it t}-test was performed on the mean RMS errors for GESI and HASPIv1 in each HL condition.
Significant differences were found for six of the sixteen combinations.
%
%
%
%
The bottom two rows show the correlation coefficients between the subjective and predicted SI values for all data points.
The correlation coefficients were always higher for GESI than for HASPIv1. 
In particular, the difference is greater in the remote male experiments.
This is probably because GESI used $\rho$ in Eq.~\ref{eq:GESI_Similarity} to adequately reflect the individual listening environment, while HASPIv1 does not provide such a control parameter.
As a result, GESI is more accurate than HASPIv1 in predicting the individual subjective SI values. 

\section{Conclusions}

In this paper, we described a new OIM called GESI that can predict the SI of simulated HL sounds for NH listeners.
GESI is an intrusive method that computes the SI metric using GCFB, modulation filterbank, and extended cosine similarity measure. 
By using GCFB, it is possible to reflect the HL that appears on the audiogram in HI listeners caused by active and passive cochlear dysfunction. GESI provides a single goodness metric that can be used immediately to evaluate SE algorithms. The metric is derived solely from the input signals without the use of training data to map internal parameters to the SI. In this sense, it is an extension of STOI and ESTOI, which are widely used to evaluate SE algorithms, to SI prediction of HI listeners who may benefit from SE algorithms in hearing assistive devices.
GESI also provides a simple control parameter that accepts the level asymmetry of the reference and test sounds and incorporates the result of the tone pip test, which can be used to estimate the participant's listening condition as determined by the sound presentation level, ambient noise level, audio equipment quality, and hearing level.

To evaluate GESI and conventional OIMs (STOI, ESTOI, MBSTOI, HASPIv1, and HASPIv2), we conducted four SI experiments on male and female word sounds in both laboratory and remote environments.
Three analyses were performed. {\it i)} Prediction of mean SI using simulated HL sounds;
{\it ii)} Prediction of mean SI without the use of simulated HL sounds with limited parameter value settings; {\it iii)} Prediction of individual listener SI.
GESI predicted word SI values better than the conventional OIMs in these evaluations. 
STOI, ESTOI, and MBSTOI did not predict SI at all, even when using the simulated HL sounds. It was also found that the NN output of HASPIv2, which is aimed at predicting sentence SI, was not suitable for predicting word SI in this evaluation. Although HASPIv1 was able to predict SI without using the simulated HL sounds, HASPIv1 did not predict well the differences between the lab and remote trials for male speech sounds and between male and female speech sounds. In addition, HASPIv1 and HASPIv2 require training data for the mapping function in order to derive a single metric for evaluating SE algorithms.

Although the current results were limited to the SI prediction for the word experiments with simulated HL sounds for NH listeners, GESI could be used to improve SE algorithms in hearing assistive devices for individual HI listeners whose HL is caused solely by peripheral dysfunction. 
GESI is available from our GitHub repository~\citep{GitHub_amlab}.

\section*{Acknowledgments}
This research was supported by JSPS KAKENHI: Grant Numbers JP21H03468 and JP21K19794.


\bibliography{Reference_Jul23}

\appendix 

\section{Pre-screening for crowdsourced remote experiments}
\label{secApndx:Pre-screening}

We conducted a pre-screening experiment before the remote experiments of the female speech sounds. 

\subsection{Motivation of the pre-screening experiment}
\label{secApndx:MotovationPre-screening}

We first performed the SI experiments on the male speech sounds in the laboratory and remote environments.
As described in section \ref{sec:SbjSI_Result} and shown in Fig.~\ref{fig:RsltSbj}(b), there was a large variability in the SI values, particularly in the low-level conditions. We assumed that this result was caused by the different listening conditions. It was also found that there was a relationship between the reported number of pips and the SRT as shown in Fig.~\ref{fig:Scatter_NpipSRT}(b).
This observation led us to perform the pre-screening experiment before the remote experiments on the female speech sounds, with the aim of reducing variability and experimental cost. We created another set of web pages for the pre-screening experiment, which can be completed in approximately 15 minutes. 


\subsection{Contents of the pre-screening experiment}
Participants first register basic information such as user ID in the crowdsourcing service, gender, and age. They are then asked to register information about the devices they use: the manufacturer, model number, and URL of wired headphones or earphones (to prevent the use of cheap disposable devices); the type of personal computer, Windows or Mac; and the external audio interface, if used. They were also asked to self-report their own HL. 

\paragraph{Volume setting}

After repeatedly confirming that they are using either wired headphones or wired earphones, participants are asked to adjust the volume to a slightly louder but comfortable level by listening to speech at the same level as the unprocessed speech used in the SI experiments.
Participants are then asked to confirm that they are able to hear the content of the speech sounds in the low-level condition. If they have difficulty, they are instructed to go back and adjust the volume.



\paragraph{Tone pip test}
The tone pip test in the male speech experiment only used 
the descending series of tone pips as shown in Fig.~\ref{fig:TonePip}.
However, an ascending series was also added for more accurate estimation.
Participants were asked to answer a total of eight questions in order to listen to descending and ascending series for the four frequencies(500\,Hz, 1000\,Hz, 2000\,Hz, and 4000\,Hz).


\paragraph{Huggins pitch test}
Huggins pitch test is conducted to confirm whether participants use headphones or earphones \citep{milne2021online}.
First, a broadband noise (white noise) is prepared. The left channel outputs the unprocessed noise as it is, and the right channel outputs the noise whose phase is changed only in the frequency band of 600 Hz (±6\%).
However, when listened to dichotically with headphones or earphones, a pitch as high as 600\,Hz can be heard in the noise, known as Huggins Pitch (HP). 
Participants are asked to respond to the interval containing HP in a three-alternative forced-choice task. Six sets are administered and the number of correct responses is calculated.



\paragraph{Vocabulary estimation test}
Vocabulary size in the mental lexicon might also affect the results because the current SI experiments are conducted with words with low familiarity. Participants take a test to estimate Japanese vocabulary size \citep{kondo2013hyakurakan}. After analyzing the results, it was found that there was no correlation between the vocabulary size and SRT in any of the current experiments.

\subsection{Criteria for pre-screening}

This pre-screening experiment was conducted twice before the crowdsourced remote experiments of female speech sounds on a first-come, first-served basis, as described in section \ref{sec:SbjSI_Remote}.
The total number of participants was 300: 
100 participants aged 21 to 63 years in the first experiment and 200 participants aged 20 to 71 years in the second experiment.
The criteria for participation in the main experiment were 
{\it i)} the average number of tone pips heard was less than and equal to 13, to eliminate those who misunderstood the task; {\it ii)} the results of the Huggins pitch tests were perfect, as was the criterion in the previous study ~\cite{milne2021online}; {\it iii)} the registered headphones or earphones were a reliable product, to avoid cheap disposable ones.
As a result, a total of 116 participants were selected for the SI experiment: 35 out of 100 for the first session and 81 out of 200 for the second session.
We asked 95 people from this pre-screened subject pool to participate in the female speech experiments.

\color{black}

\end{document}